\documentclass[twocolumn]{aastex63}
\usepackage{natbib}
\usepackage{amsmath}
\usepackage{chngcntr}
\usepackage{hyperref}
\usepackage[figuresright]{rotating}
\usepackage{ctable}
\usepackage{graphicx}
\usepackage{realboxes}
\usepackage{epsfig}
\usepackage{threeparttablex, tablefootnote}
\usepackage{subfigure}
\usepackage{graphicx}
\newcommand{\uat}[2]{\href{http://astrothesaurus.org/uat/#2}{#1 (#2)}}
\usepackage{CJK}
\shortauthors{Han et al.}

\begin{document}

\begin{CJK*}{UTF8}{gbsn}
\title{Stellar chromospheric activities revealed from the LAMOST-\emph{K2} time-domain survey}

\correspondingauthor{Song Wang, Henggeng Han}
\email{songw@bao.ac.cn, hghan@nao.cas.cn}

\author{Henggeng Han}
\affil{Key Laboratory of Optical Astronomy, National Astronomical Observatories, Chinese Academy of Sciences, Beijing 100101, China}
\affil{Institut f\"ur Physik und Astronomie, Universit\"at Potsdam, Karl-Liebknecht-Str. 24/25, 14476 Potsdam, Germany}
\affil{School of Astronomy and Space Science, University of Chinese Academy of Sciences, Beijing 100049, China}

\author{Song Wang}
\affil{Key Laboratory of Optical Astronomy, National Astronomical Observatories, Chinese Academy of Sciences, Beijing 100101, China}

\author{Yu Bai}
\affil{Key Laboratory of Optical Astronomy, National Astronomical Observatories, Chinese Academy of Sciences, Beijing 100101, China}

\author{Huiqin Yang}
\affil{Key Laboratory of Optical Astronomy, National Astronomical Observatories, Chinese Academy of Sciences, Beijing 100101, China}

\author{Xiangsong Fang}
\affil{Key Laboratory of Optical Astronomy, National Astronomical Observatories, Chinese Academy of Sciences, Beijing 100101, China}

\author{Jifeng Liu}
\affil{Key Laboratory of Optical Astronomy, National Astronomical Observatories, Chinese Academy of Sciences, Beijing 100101, China}
\affil{School of Astronomy and Space Science, University of Chinese Academy of Sciences, Beijing 100049, China}
\affil{WHU-NAOC Joint Center for Astronomy, Wuhan University, Wuhan, Hubei 430072, China}

\begin{abstract}

By using the LAMOST time-domain survey data, we study stellar activities based on the $\rm{H_{\alpha}}$ lines for about 2000 stars in four $K$2 plates. 
Two indices, $R_{\rm{H\alpha}}^{'}$ and $R_{\rm{H\alpha}}^{+}$, are computed from LAMOST spectra, the former of which is derived by excluding the photospheric contributions to the $\rm{H_{\alpha}}$ lines, while the latter is derived by further subtracting the non-dynamo driven chromospheric emission. Meanwhile, the periodicity and variation amplitudes are computed from \emph{K2} light curves.
Both the $R_{\rm{H\alpha}}^{'}$-Ro relation and $R_{\rm{H\alpha}}^{+}$-Ro relation show complicated profiles in the non-saturated decay region.
Hot stars show flatter slopes and higher activity level than cool stars, and the behaviour is more notable in the $R_{\rm{H\alpha}}^{+}$-$R_{o}$ relation.
This is consistent with recent studies using other activity proxies, including $L_{\rm{x}}/L_{\rm{bol}}$, $R_{\rm{HK}}^{'}$ and amplitudes of optical light curves. %
This may suggest different kinds of stars follow different power laws in the decay region.
Most of our targets have multiple observations, and some of them exhibit significant variability of ${\rm{H\alpha}}$ emissions, which may cause the large scatters shown in the decay region.
We find three targets exhibiting positive correlation in rotational phase, possibly indicating that their optical light curves are dominated by hot faculae rather than cool starspots.

\end{abstract}

\section{Introduction} 

Stellar magnetic fields are the energy source of stellar activities. According to the dynamo theories, magnetic fields are generated by differential rotation in deep convection zones \citep{1955ApJ...122..293P, 1955ApJ...121..491P, 1984ApJ...279..763N, 2006A&A...446.1027C} or interaction of flow turbulence \citep{1993SoPh..145..207D,1996ApJ...469..828D}. It has been widely accepted that strength of magnetic fields is positively related to stellar activity level.

The strength of magnetic activity can be traced by various proxies, including spots, flares, chromospheric emissions, X-ray and radio emissions. The golden one is the Ca \scriptsize{\uppercase\expandafter{\romannumeral2}} \normalsize H$\&$K emission lines, whose cores are extremely sensitive to magnetic fields \citep{1955ApJ...121..349B, 1959ApJ...130..366L}. \cite{1978ApJ...226..379W} carried out long-term observations of chromospheric activities for different types of stars. In order to quantify the activity level, \cite{1978PASP...90..267V} introduced the well known S-index. Later on, \cite{1984ApJ...279..763N} proposed $R_{\rm{HK}}^{'}$ value, which can characterize the excess chromospheric emission of Ca \scriptsize{\uppercase\expandafter{\romannumeral2}} \normalsize H$\&$K lines.


However, for faint cool stars the Ca \scriptsize{\uppercase\expandafter{\romannumeral2}} \normalsize H$\&$K lines are less notable compared to those of hot stars so that it is hard to detect. An alternative proxy of chromospheric activity is the $\rm{H_{\alpha}}$ emission. Although such emission could be dominated by photoionization, as the temperature decreases the contribution of collisional excitation would gradually become significant and thus the $\rm{H_{\alpha}}$ line can be used as tracer of chromorpheric activities \citep{2007A&A...469..309C, 2017ARA&A..55..159L}. 

Chromospheric activity levels are always related to stellar rotation. The well known activity-rotation relation provides fundamental information on stellar dynamos and angular momentum evolution, and has been comprehensively studied. There are different proxies to trace this relation, including X-ray, $\rm{H_{\alpha}}$ line and Ca \scriptsize{\uppercase\expandafter{\romannumeral2}} \normalsize H$\&$K lines. For different tracers, the activity-rotation relations show similar trend, e.g., a flat saturated region and a power-law decay region against rotation periods or Rossby number (Ro) in logarithmic scale \citep[e.g.][]{2003A&A...397..147P,2011ApJ...743...48W}. 

Such standard relation has been challenged by many recent studies. \citet{2019A&A...628A..41P} found that instead of continuous decaying, some X-ray emitting $Kepler$ stars with Ro $>$ 0.3 behaved differently in the such relation (See their Figure 11). \cite{2021ApJ...910..110L} argued that such relation should consist of two pieces of power-law. \citet{2018A&A...618A..48M} even divided the relation into four regions 
after combining the stellar X-ray and Ca \scriptsize{\uppercase\expandafter{\romannumeral2}} \normalsize H$\&$K emission. By calculating the photometric variability ($R_{\rm var}$) for tens of thousands of $K$2 stars, \citet{2020A&A...635A..43R} presented the relation between $R_{\rm var}$ and rotation periods of different types of stars. Interestingly, a rather flat relation was shown when periods are longer than 20 days.

Motivated by these new findings, in this study, 
we adopted the $\rm{H_{\alpha}}$ emission to investigate the relation between magnetic activity and rotation for various types of stars based on the recent LAMOST time resolved sky survey and \emph{K2} light curves. An ideal sample to study the connection between different activity proxies and mechanism of stellar dynamos is also provided. The paper is organized as follows.
In Section 2 we introduce the sample and data reduction. 
Section 3 shows the main results, including relative corrections of equivalent widths, the calculation of normalized $\rm{H_{\alpha}}$ luminosities, and the estimation of rotation periods and Rossby numbers.
We discuss our results in Section 4, including the activity-rotation relation based on $\rm{H_{\alpha}}$ emission, and the relation between $\rm{H_{\alpha}}$ emission and $K$2 brightness in the rotational phase.


\begin{figure}
\centering
\includegraphics[width=0.5\textwidth]{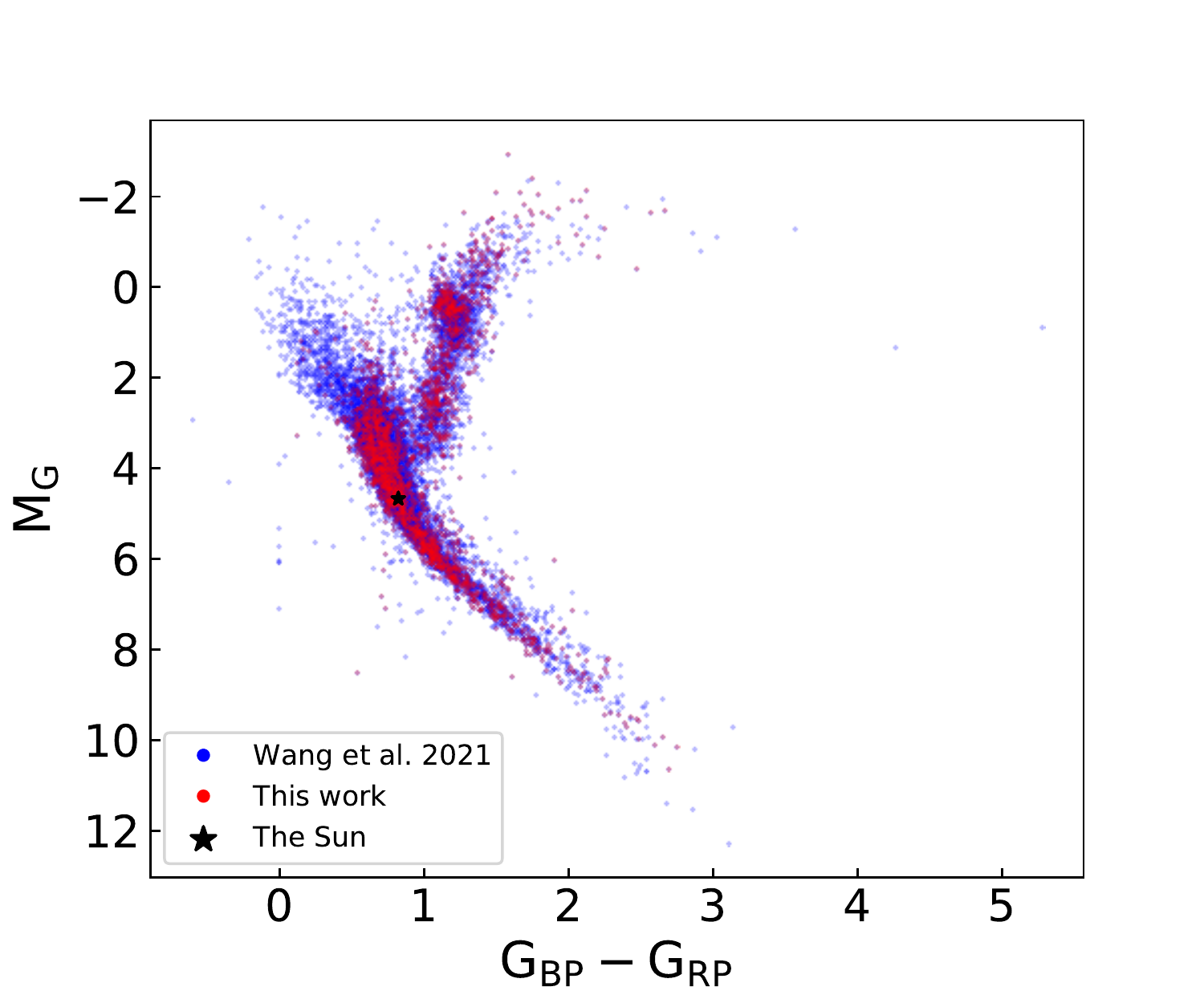}
\caption{Colour-magnitude diagram. Sample of \cite{2021RAA....21..292W} and our targets are shown in blue and red dots, respectively. Black star represents the Sun.}
\label{HRD.fig}
\end{figure}

\section{Source Selection and Data Reduction}
\subsection{Source selection}
\label{sel.sec}

\begin{figure}
\centering
\includegraphics[width=0.45\textwidth]{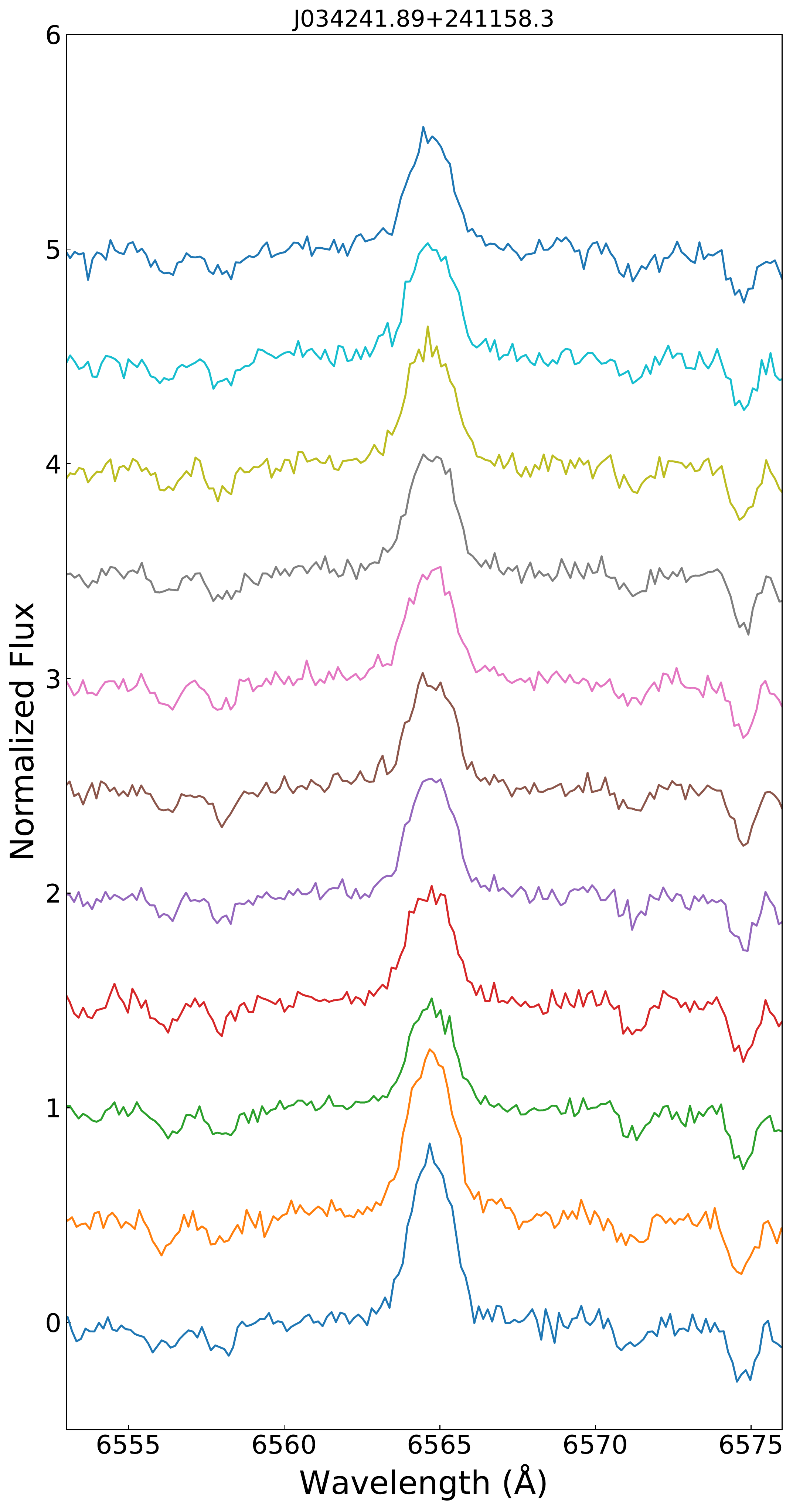}
\caption{Examples of $\rm{H_{\alpha}}$ emission lines of J034241.89+241158.3. The spectra are shifted for clarity.}
\label{example.fig}
\end{figure}

The \emph{Kepler} mission and its extended \emph{K2} mission provided light curves with high photometric precision of more than 200,\,000 stars. Such a huge and elite sample would open a new era of studying stellar physics \citep{2010Sci...327..977B, 2010ApJ...713L..79K}. The Large Sky Area Multi-Object Fibre Spectroscopic Telescope (hereafter LAMOST; also named as Guoshoujing Telescope) is a quasi-meridian Schmidt telescope with a 4-meter aperture and a field of view of 5 degree \citep[LAMOST;][]{2012RAA....12.1197C}. With 4000 fibers on its focus surface, tens of millions of spectra have been gathered with high efficiency. 

Since 2012, the LAMOST-\emph{Kepler/K2} projects have been carried out, which performed both time-domain and non time-domain sky survey of the \emph{Kepler} field and the \emph{K2} campaigns, releasing both the medium-resolution spectra(MRS) and low-resolution spectral(LRS) data for tens of thousands of targets \citep{2020RAA....20..167F, 2020ApJS..251...15Z}. Recently, LAMOST has started the second 5-year sky survey, which performs both LRS and MRS observations with $\Delta \lambda / \lambda \sim$ 1800 and $\sim$ 7500, respectively \citep{2020arXiv200507210L, 2020ApJS..251...15Z}. The MRS observations provide spectra covering the wavelength range from 4950 \AA \ to 5930 \AA \ for the blue arm and from 6300 \AA \ to 6800 \AA \ for the red arm, respectively. Spectra of the LRS observations cover the wavelength range of 3650--9000 \AA. 

Recently, \cite{2021RAA....21..292W} reported the first results of the LAMOST time-domain survey, covering four \emph{K2 plates}. There are 10,700 targets in their sample and most of them have multiple observations. For the LRS observations, 767,158 spectra were derived in blue arms and 767,150 spectra in red arms. For the MRS observations, 478,694 spectra were gathered for both the blue and red arms. 

Among these targets, over 3000 targets have \emph{K2} light curves. In \cite{2021RAA....21..292W}, the \emph{Lomb-Scargle} method \citep{1976Ap&SS..39..447L, 1982ApJ...263..835S} was applied for period detection. In brief, a two-step grid searching method was carried out to determine the optimized period \citep{2015ApJ...812...18V}, including searching in a broad grid for a series of period candidates and zooming in on a narrow grid to find the real peak. All the light curves were folded based on the detected periods and the variable types of stars were classified by using light curve templates \citep[e.g.,][]{2014A&A...566A..43K}.
These objects are our input sample.

In order to exclude potential binaries, we calculated the radial velocity (RV) for each object based on multiple MRS exposures through the cross-correlation method.
The \emph{PHOENIX} high-resolution synthetic spectra \citep{2013A&A...553A...6H} were used as templates and were convolved to the MRS resolution ($R$ = 7500).
The cross-correlation results were visually checked to exclude double-lined spectroscopic binaries. 
Meanwhile, targets with RV variation larger than 10 km/s were also removed. In addition, we excluded possible pulsating variables and potential other types of variables given in \cite{2021RAA....21..292W}. This yield 2454 stars. 

Figure \ref{HRD.fig} plots the colour-magnitude diagram of both the \cite{2021RAA....21..292W} sample and our targets in blue and red dots, respectively. Our sample contains stars with different spectral types and there are 1856 dwarfs and 593 giants, among which there are 17 A-type stars, 505 F-type stars, 684 G-type stars, 1140 K-type stars and 103 M-type stars. 
Stellar parameters, including the effective temperature ($T_{\rm eff}$), surface gravity (log$g$) and metallicity [Fe/H], of these stars were extracted from LAMOST DR8 catalog, which were estimated base on the LAMOST Stellar Parameter Pipline \citep{2015RAA....15.1095L}.


\subsection{Equivalent Width measurements} 
\label{ew.sec}
Only the spectra with signal-to-noise ratio (SNR) higher than 10 were reserved for our spectral analysis. These spectra were normalized for further equivalent widths calculation\citep{2020ApJS..246....9Z, 2021ApJS..256...14Z}. Figure \ref{example.fig} shows an example of $\rm{H\alpha}$ emission of target J034241.89+241158.3. Generally, there are two ways to calculate the equivalent widths: integration and Gaussian fitting.
Widths of Balmer lines would decrease with effective temperatures \citep{1978stat.book.....M}. 
Therefore, the wavelength range for the calculation should be different for each type of stars. We followed four steps to test the results by changing the wavelength range. Detailed steps are listed as follows,

\begin{figure*}
\centering
\subfigure[]{
\includegraphics[width=0.45\textwidth]{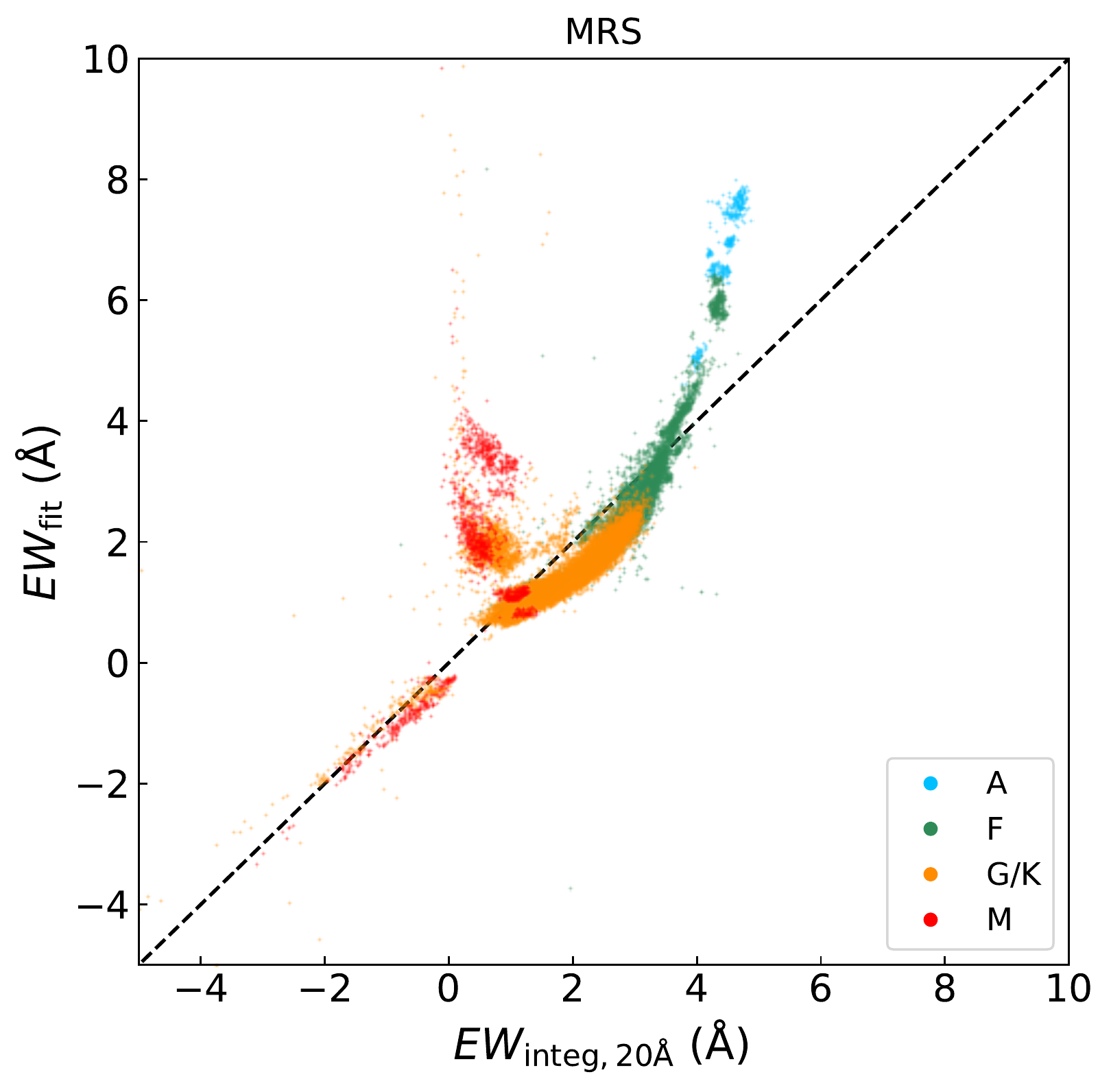}}
\subfigure[]{
\includegraphics[width=0.45\textwidth]{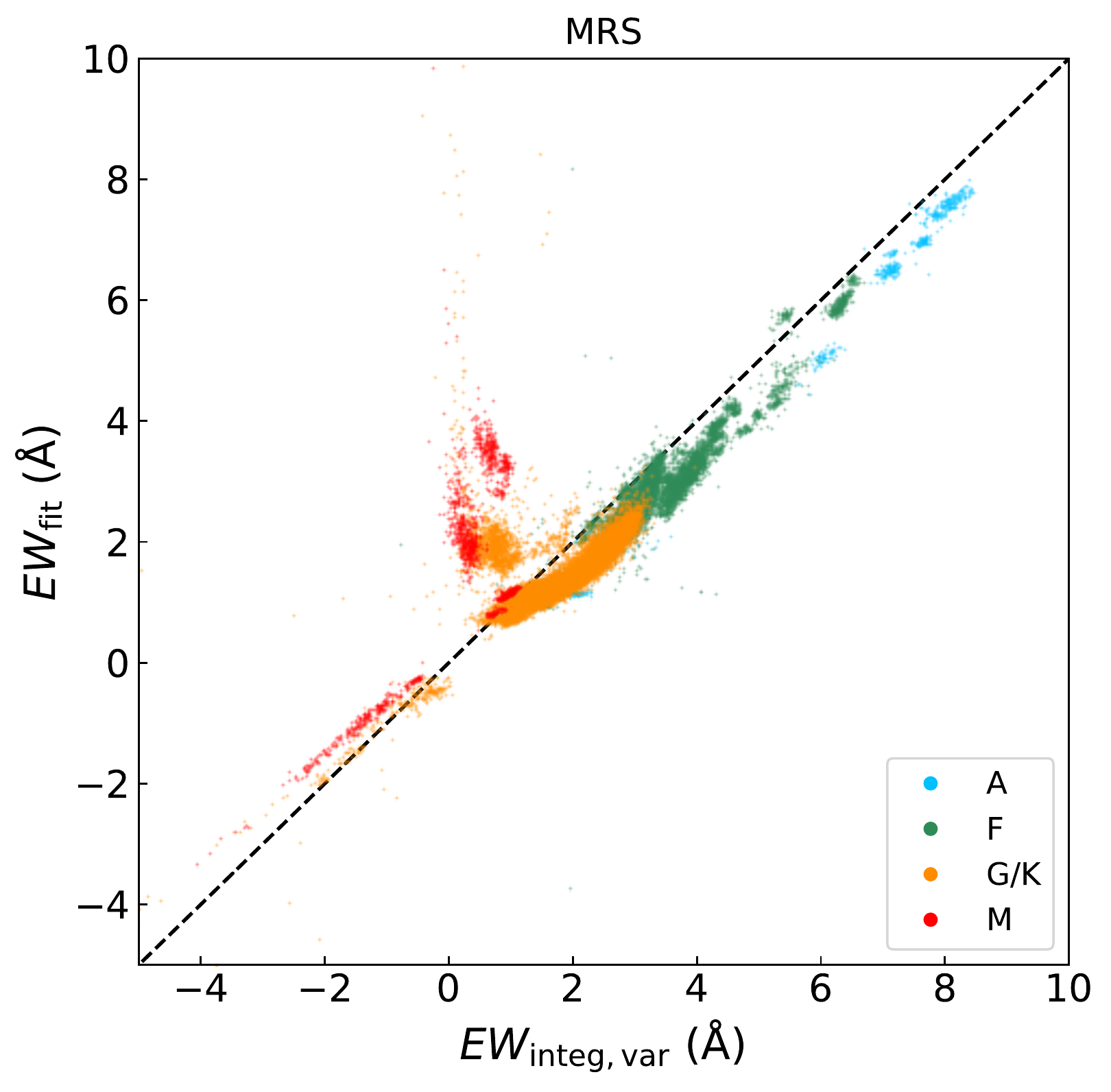}}
\subfigure[]{
\includegraphics[width=0.45\textwidth]{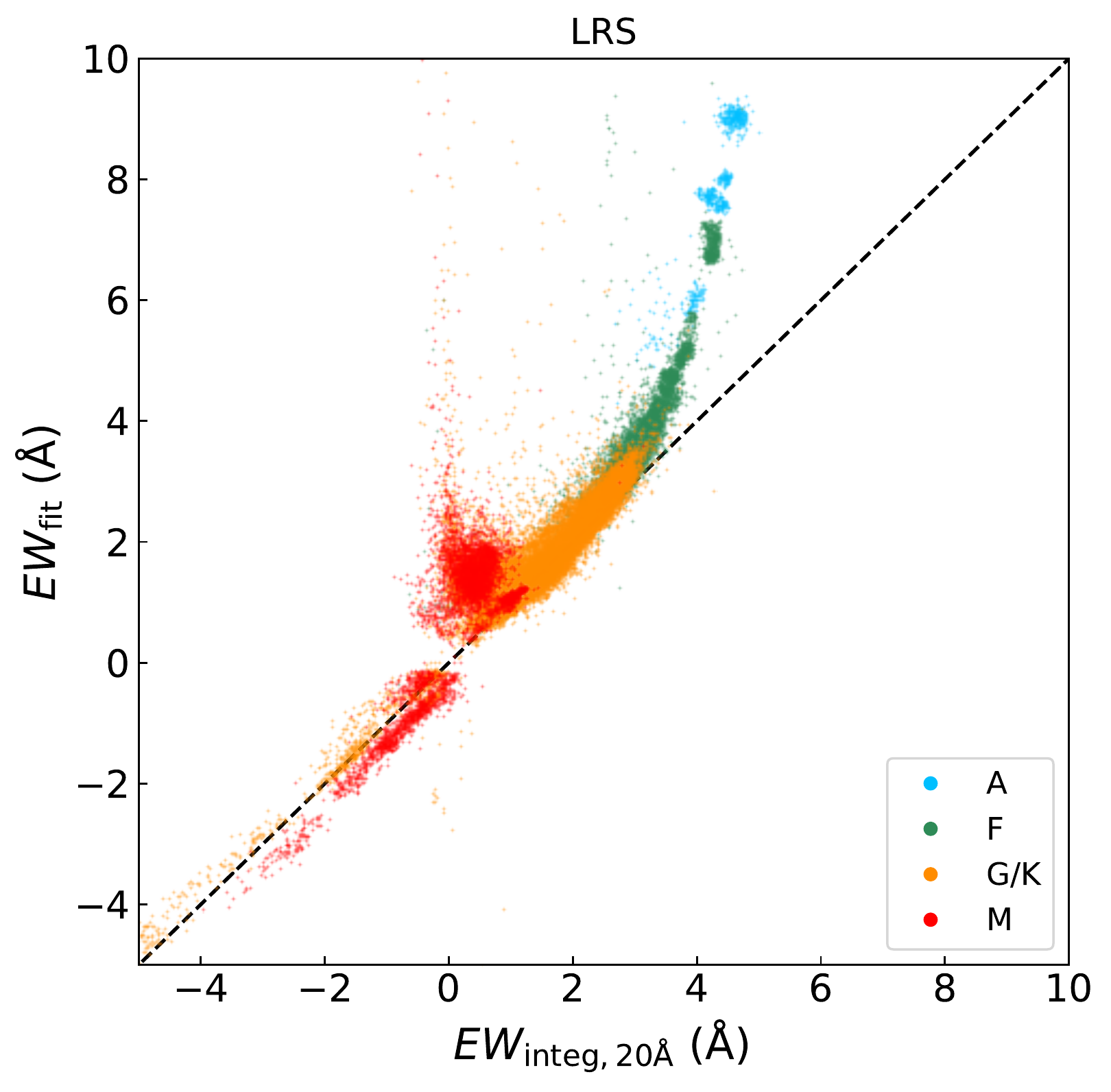}}
\subfigure[]{
\includegraphics[width=0.45\textwidth]{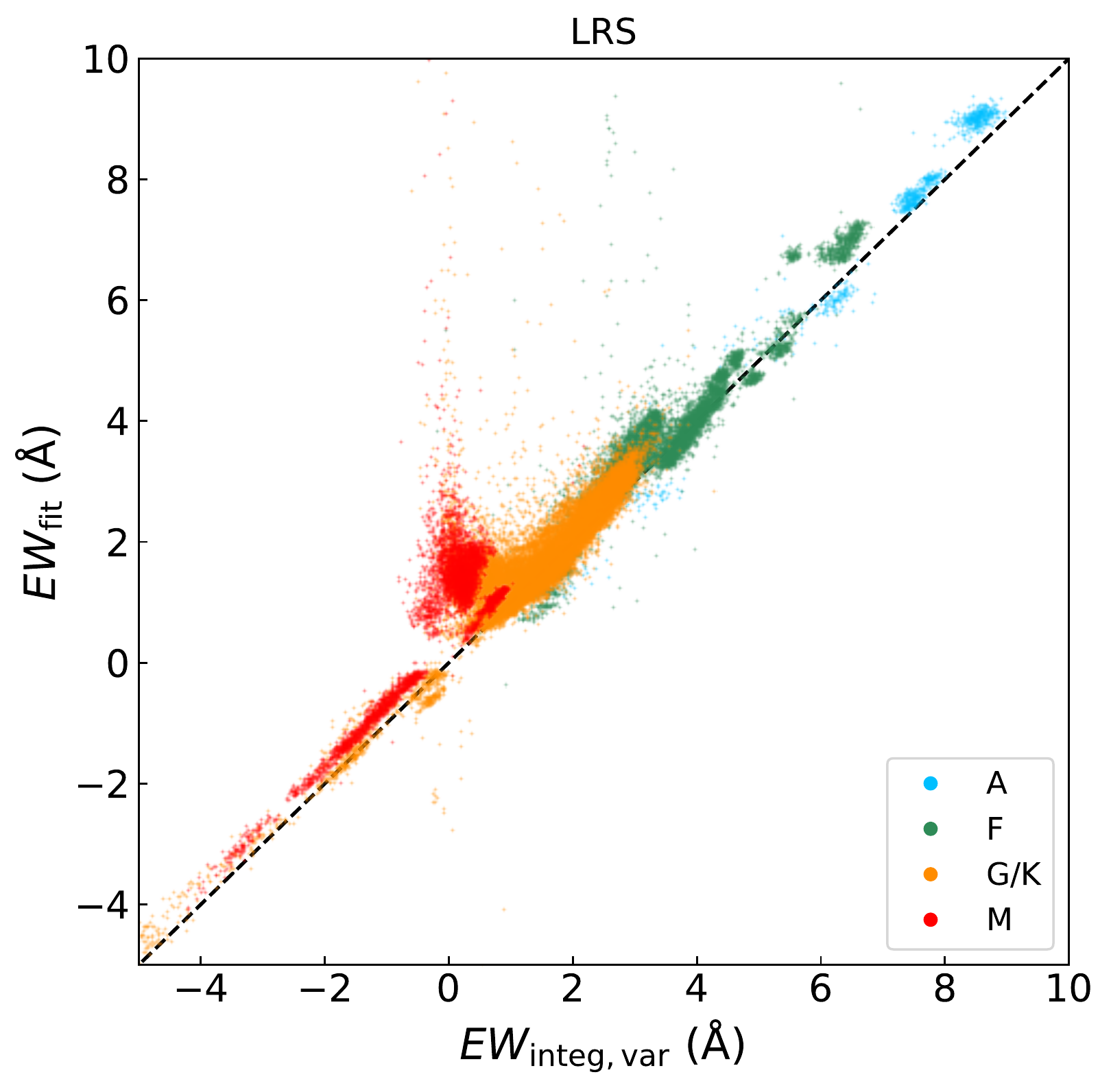}}
\caption{Comparison between Gaussian fitting and integration method for calculating the $EWs$ of $\rm{H\alpha}$ lines. Panel (a) and panel (b) are results of MRS observations. Panel (c) and panel (d) are results of LRS observations. Different colours represent different stellar types.}
\label{comgauss.fig}
\end{figure*}

\begin{figure*}
\centering
\subfigure[]{
\includegraphics[width=0.45\textwidth]{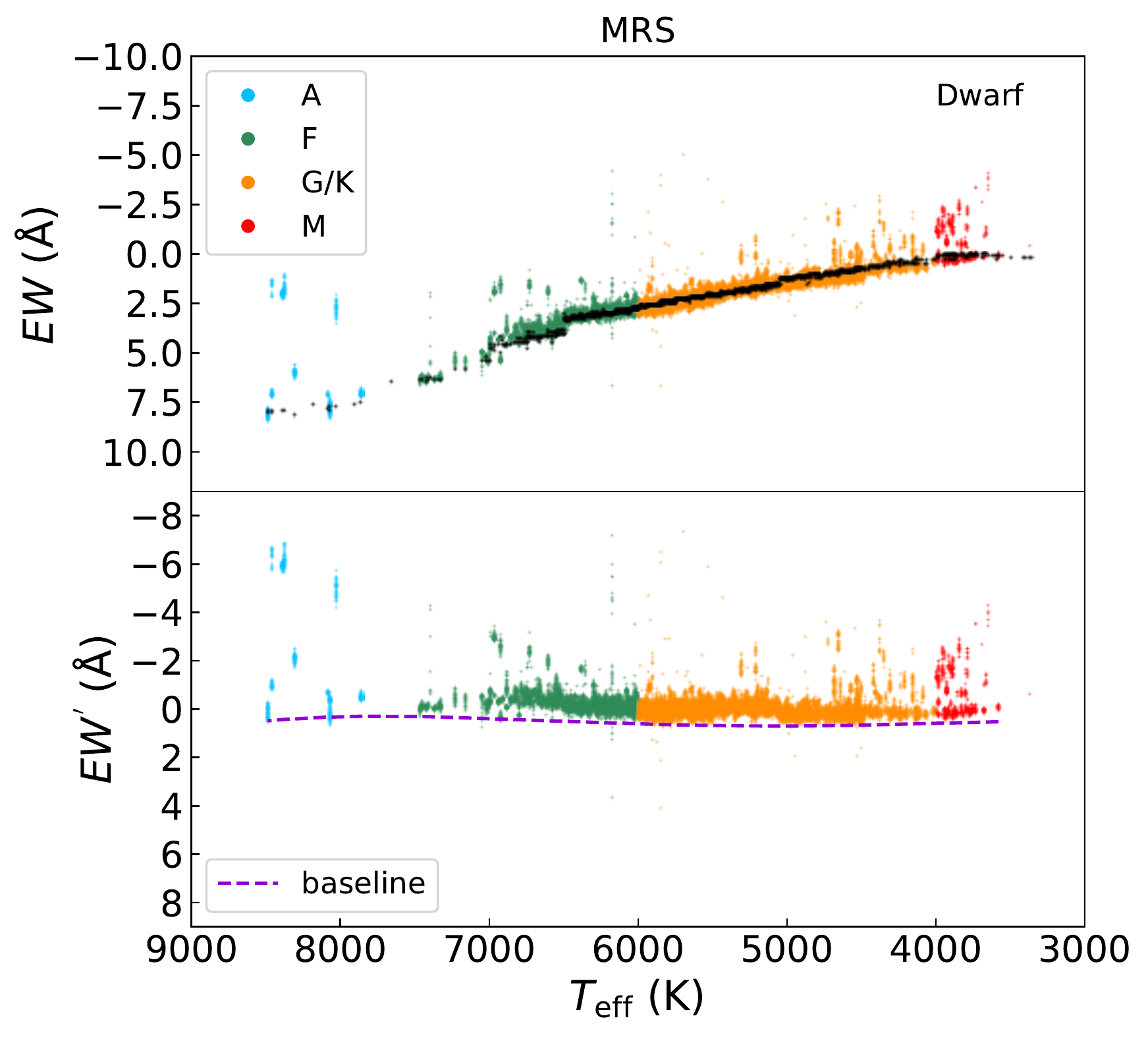}}
\subfigure[]{
\includegraphics[width=0.45\textwidth]{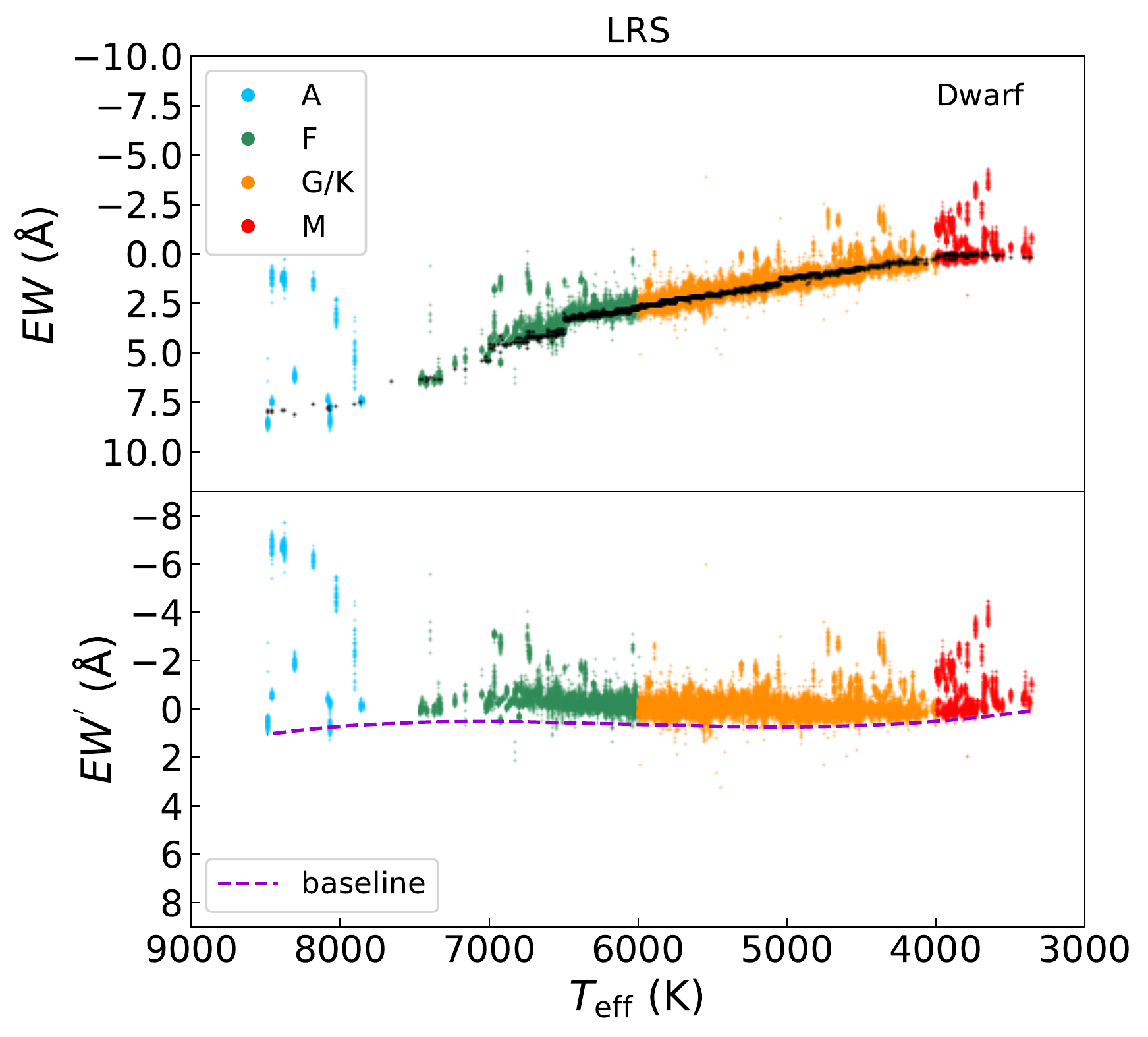}}
\subfigure[]{
\includegraphics[width=0.45\textwidth]{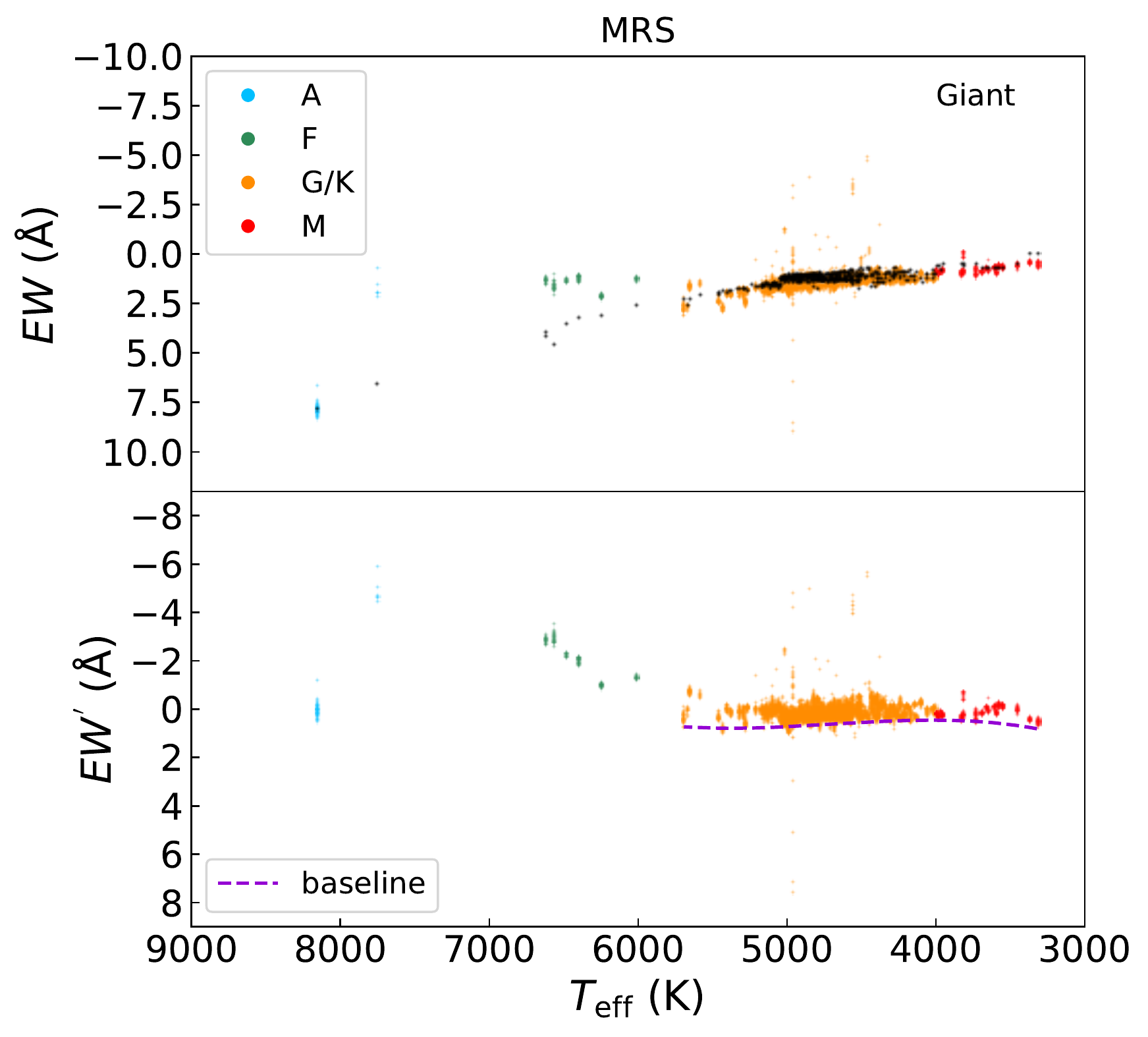}}
\subfigure[]{
\includegraphics[width=0.45\textwidth]{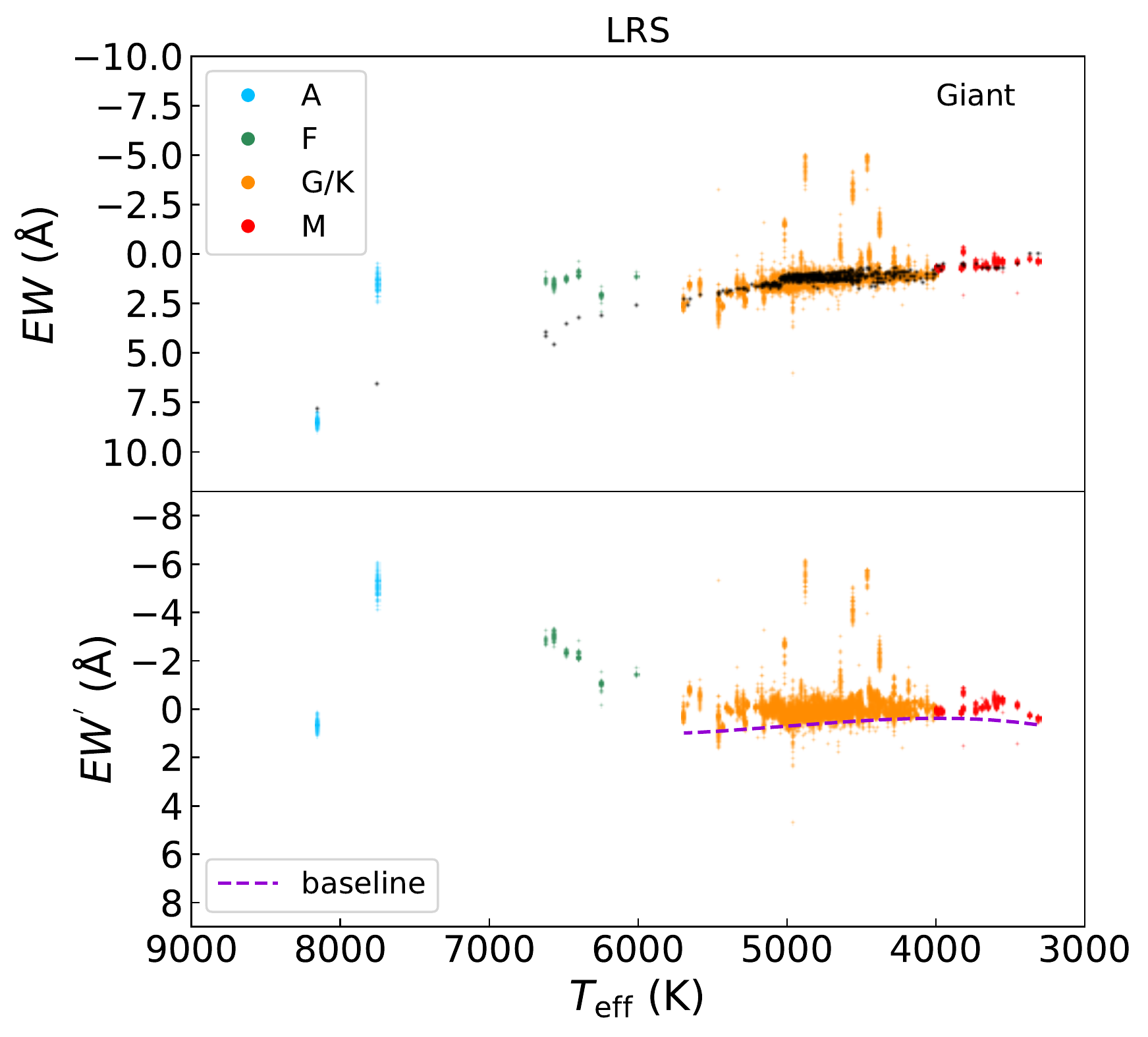}}
\caption{$EWs$ of $\rm{H_{\alpha}}$ lines versus stellar effective temperatures. 
Panel (a) and panel (b) are results of dwarf stars, but for MRS and LRS data, respectively.
Panel (c) and panel (d) are results of giant stars, but for MRS and LRS data, respectively.
Points in different colours represent different stellar types. 
Black dots are photospheric contributions of $\rm{H_{\alpha}}$ lines calculated from the \emph{PHOENIX} models. Dashed purple lines are baselines.}
\label{basal.fig}
\end{figure*}

\begin{table}[t]
  \begin{center}
    \caption{MRS $EWs$ of the $\rm{H{\alpha}}$ lines.}
    \label{tab:table1}
    \begin{tabular}{ccc}
    \hline\noalign{\smallskip}
      ID & EW  & BMJD  \\
       &  (\AA) & (day) \\
        (1) & (2) & (3) \\
      \hline  
J035106.15+222205.6 & 1.7 $\pm$ 0.08 & 58801.7\\
J035106.15+222205.6 & 1.55 $\pm$ 0.07 & 58801.72\\
J035106.15+222205.6 & 1.72 $\pm$ 0.08 & 58801.74\\
J035106.15+222205.6 & 1.51 $\pm$ 0.09 & 58801.75\\
J035106.15+222205.6 & 1.4 $\pm$ 0.07 & 58819.62\\
J035106.15+222205.6 & 1.67 $\pm$ 0.07 & 58819.63\\
J035106.15+222205.6 & 1.55 $\pm$ 0.07 & 58819.64\\
J035106.15+222205.6 & 1.52 $\pm$ 0.09 & 58819.65\\
J035106.15+222205.6 & 1.59 $\pm$ 0.06 & 58819.66\\
J035106.15+222205.6 & 1.52 $\pm$ 0.07 & 58819.67\\
J035106.15+222205.6 & 1.59 $\pm$ 0.07 & 58890.45\\
J035106.15+222205.6 & 1.66 $\pm$ 0.08 & 58890.47\\
J035106.15+222205.6 & 1.34 $\pm$ 0.1 & 58890.49\\
J035106.15+222205.6 & 1.56 $\pm$ 0.1 & 58890.5\\
... & ... & ... \\
      \hline \\
    \end{tabular}
  \end{center}
\end{table}

\begin{table}
  \begin{center}
    \caption{LRS $EWs$ of the $\rm{H{\alpha}}$ lines.}
    \label{tab:table2}
    \begin{tabular}{ccc}
    \hline\noalign{\smallskip}
      ID & EW  & BMJD  \\
       &  (\AA) & (day) \\
        (1) & (2) & (3)  \\
      \hline  
J035106.15+222205.6 & 1.5 $\pm$ 0.04 & 58784.68\\
J035106.15+222205.6 & 1.67 $\pm$ 0.04 & 58784.69\\
J035106.15+222205.6 & 1.44 $\pm$ 0.05 & 58784.7\\
J035106.15+222205.6 & 1.64 $\pm$ 0.04 & 58784.71\\
J035106.15+222205.6 & 1.53 $\pm$ 0.06 & 58784.72\\
J035106.15+222205.6 & 1.54 $\pm$ 0.06 & 58784.73\\
J035106.15+222205.6 & 1.63 $\pm$ 0.05 & 58784.74\\
J035106.15+222205.6 & 1.6 $\pm$ 0.03 & 58784.75\\
J035106.15+222205.6 & 1.48 $\pm$ 0.02 & 58787.72\\
J035106.15+222205.6 & 1.68 $\pm$ 0.02 & 58787.73\\
J035106.15+222205.6 & 1.63 $\pm$ 0.02 & 58787.74\\
J035106.15+222205.6 & 1.58 $\pm$ 0.03 & 58811.65\\
J035106.15+222205.6 & 1.61 $\pm$ 0.03 & 58811.66\\
J035106.15+222205.6 & 1.67 $\pm$ 0.04 & 58811.66\\

... &  ... & ... \\
      \hline \\
    \end{tabular}
  \end{center}
\end{table}

\begin{enumerate}
\item We divided our sample into several groups according to their effective temperatures. 
A-type stars are hotter than 7500 K and cooler than 10000 K.
Targets with effective temperatures between 6000 K to 7500 K were marked as F-type stars. 
For G-type and K-type stars, the temperature range was set to be from 5300 K to 6000 K, from 4000 K to 5300 K, respectively. 
Targets with effective temperatures lower than 4000 K were treated as M-type stars.

 \item The line centers of $\rm{H\alpha}$ were corrected using the RVs derived in Section \ref{sel.sec}.

 \item For the integration method we used the formula: 
 \begin{equation}
 EW_{\rm{H}\alpha} = \int \frac{F_{c} - F_{\lambda}}{F_{c}}d\lambda
 \end{equation}
Here $F_{c}$ is the median value of the pseudo-continua. For all kinds of stars, the range of the pseudo-continua outside the line range on both sides were set to be 10 \AA. 

 \item We tested two sets of the integration ranges: fixed range (20 \AA) and variable range. In the latter set, for A-, G-, K- and M-type stars, the widths of the line around the line center were set to be 50\AA, \, 20\AA,\, 20\AA\, and 10 \AA,\, respectively. For early F stars the $\rm{H\alpha}$ line is wider than late F stars. 
 Thus we divided F stars into three sub-samples using a temperature bin of 500 K. For stars with 7000 K to 7500 K, 6500 K to 7000 K, and 6000 K to 6500 K, the integration ranges of the $\rm{H_{\alpha}}$ lines were set to be 40\AA, \,30\AA, and 20\AA, \, respectively. 
 This step leads to two equivalent width results, named $EW_{\rm{integ, 20\AA}}$ and $EW_{\rm{integ, var}}$.

 \item In the Gaussian fitting method, we set the wavelength range as a constant, i.e., 20\AA, for all types of stars. The corresponding equivalent widths were named as $EW_{\rm fit}$.

\end{enumerate}
 
The comparisons between different methods are shown in Figure \ref{comgauss.fig}. 
It is clear that for hot stars, $EW_{\rm{integ, 20\AA}}$ significantly deviate from $EW_{\rm fit}$ due to the wide ${\rm{H\alpha}}$ profile, while $EW_{\rm{integ, var}}$ agree well with $EW_{\rm fit}$.
When the lines are weak, equivalent widths from the integration method are close to zero whereas those from Gaussian fitting would become unreliable, since those lines would deviate from a Gaussian shape and the low SNR will result in poor fittings.
Therefore, we preferred to use $EW_{\rm{integ, var}}$ (hereafter $EW$) in following analyses.

We then estimated the errors of the $EWs$. For each spectrum, 1\,000 synthetic spectra were simulated by adding Gaussian noise to each wavelength using the flux uncertainty given by LAMOST data. All the $EWs$ of these 1000 spectra were measured and the standard deviation was used as the error of $EW$. The results of the $EWs$ from MRS and LRS observations are listed in Table \ref{tab:table1} and \ref{tab:table2}, respectively.

\subsection{Correction of EWs}

In order to remove photospheric contribution to the $\rm{H_{\alpha}}$ lines, for each target, we calculated the $EW_{\rm phot}$ of $\rm{H_{\alpha}}$ absorption lines using the the \emph{PHOENIX} high-resolution spectra \citep{2013A&A...553A...6H}. Model templates were picked out based on $T_{\rm eff}$, log$g$, and [Fe/H] of our sample. The $EW_{\rm phot}$ were also calculated in the same wavelength range mentioned in Section \ref{ew.sec} according to different stellar types. 

Then the chromospheric emission in $\rm{H_{\alpha}}$ line was defined as:
\begin{equation}
EW^{'} = EW - EW_{\rm phot}.
\end{equation}
It is necessary to treat dwarfs and giants separately. In this work, stars with \rm{log}\emph{g} $\geq$ 3.5 were regarded as dwarfs and others were marked as giants. The results are shown in Figure \ref{basal.fig}

\section{Results}


\begin{figure}
\centering
\includegraphics[width=0.45\textwidth]{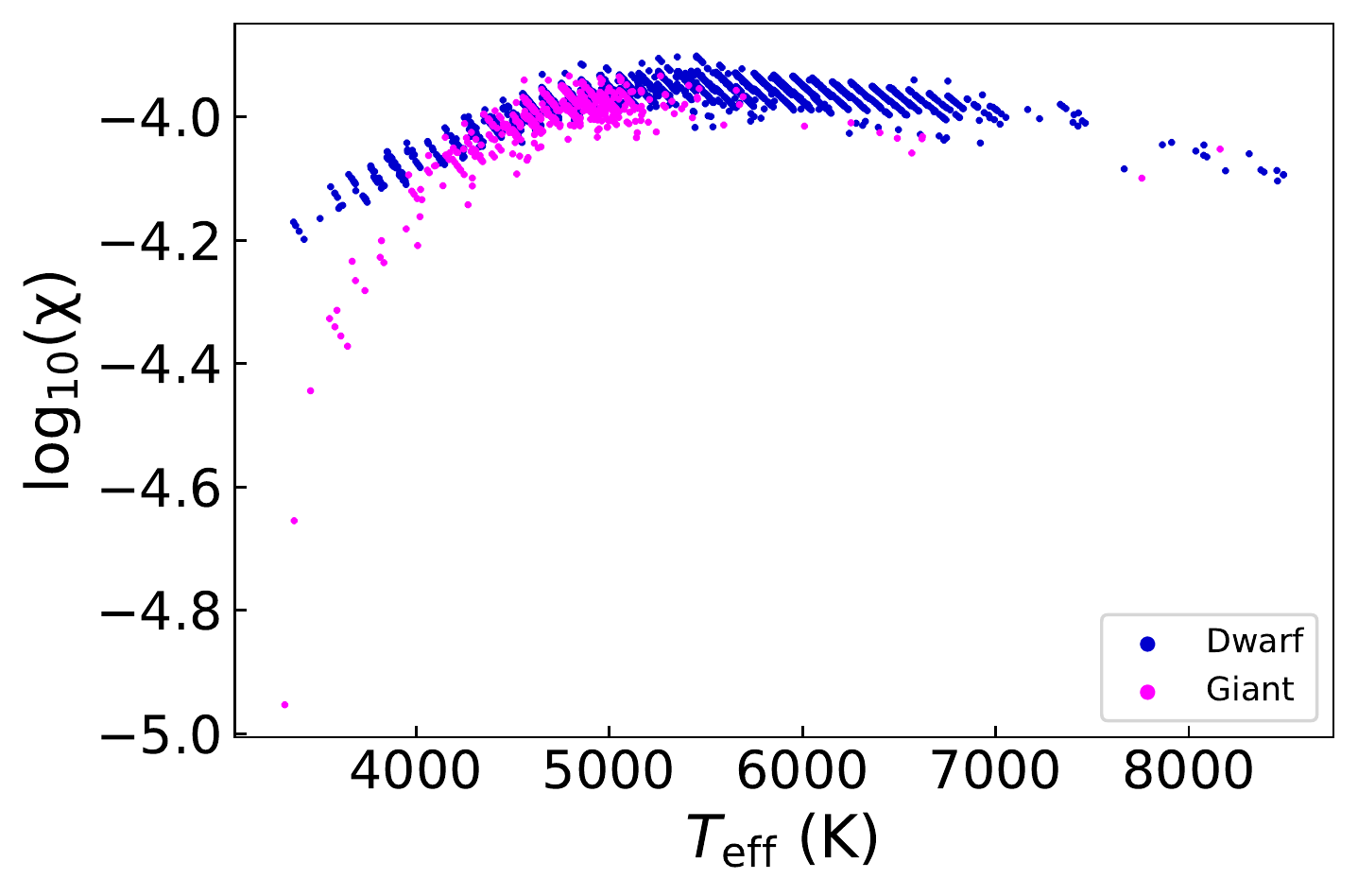}
\caption{$\chi$ versus effective temperature. Blue dots and purple dots represent dwarfs and giants, respectively.}
\label{chi.fig}
\end{figure}

\begin{figure*}
\centering
\subfigure[]{
\includegraphics[width=55ex]{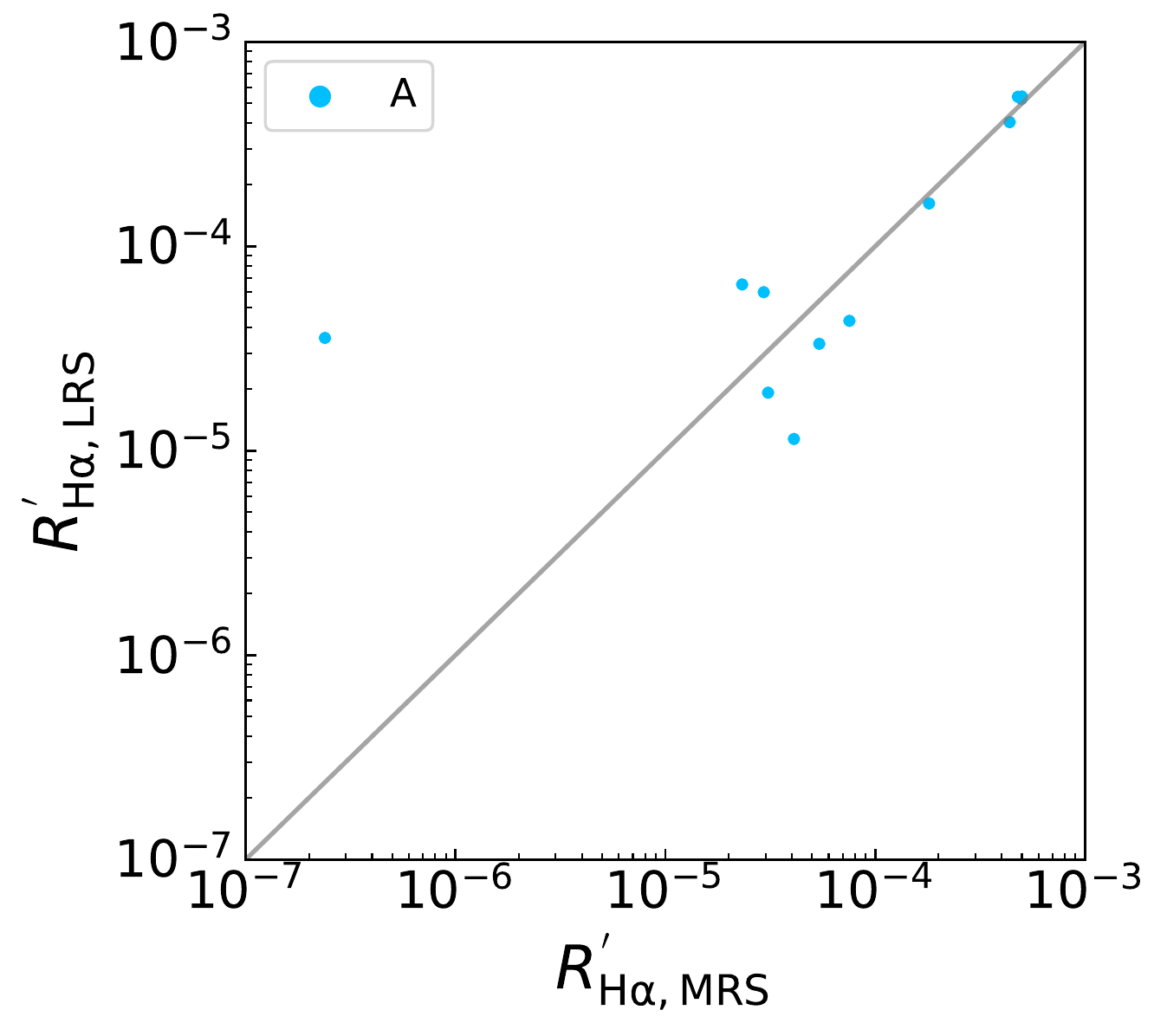}}
\subfigure[]{
\includegraphics[width=55ex]{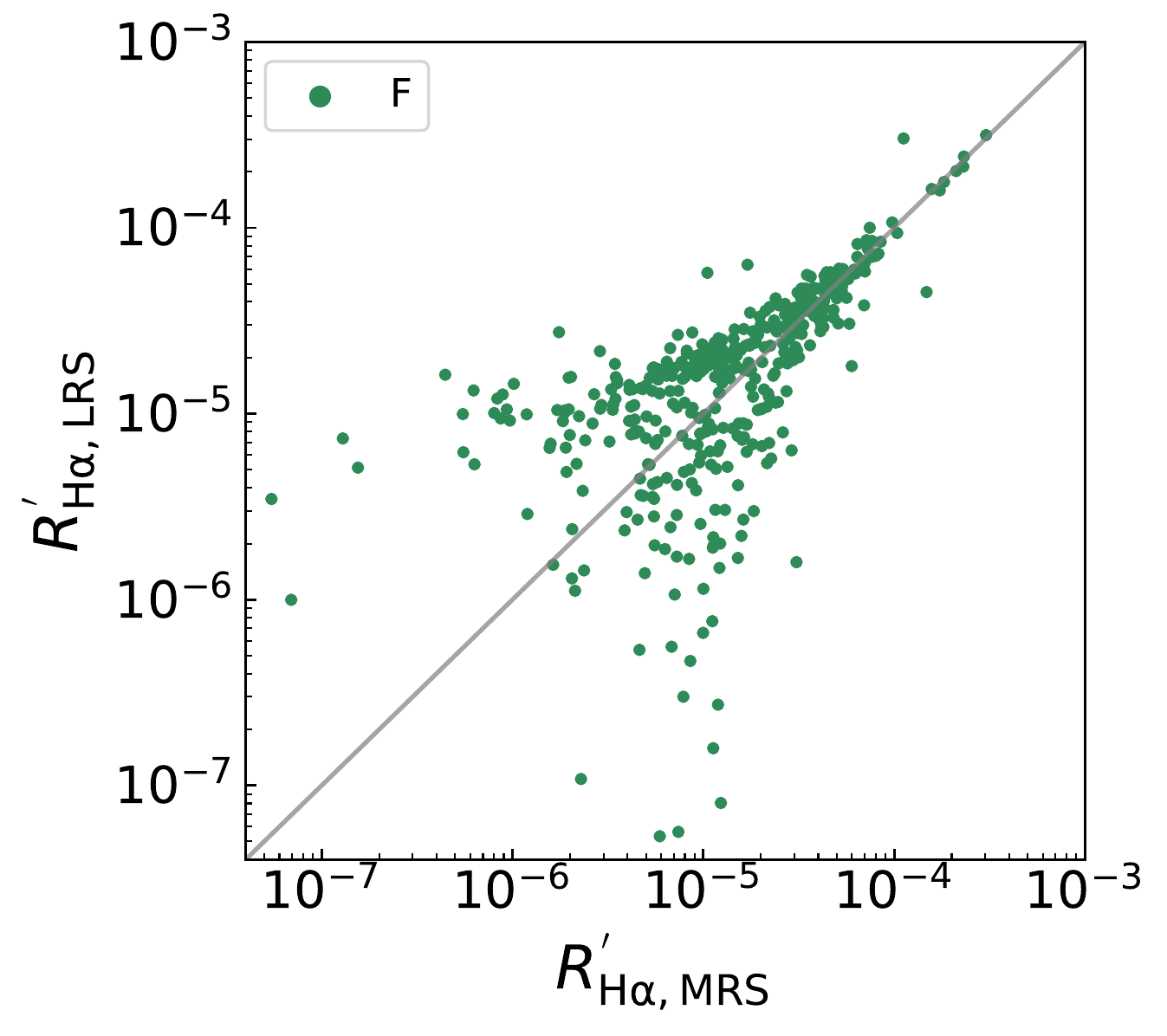}}
\subfigure[]{
\includegraphics[width=55ex]{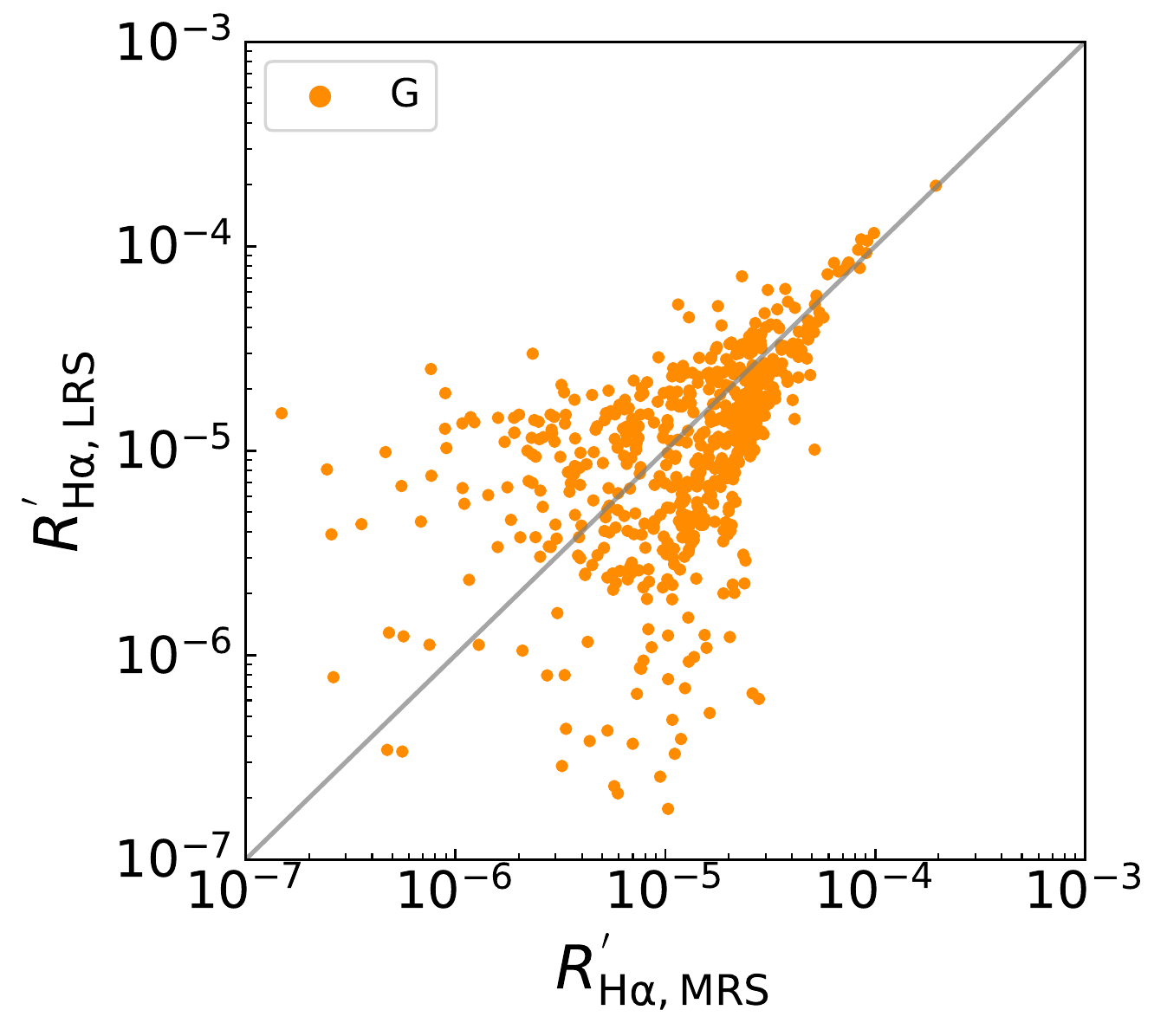}}
\subfigure[]{
\includegraphics[width=55ex]{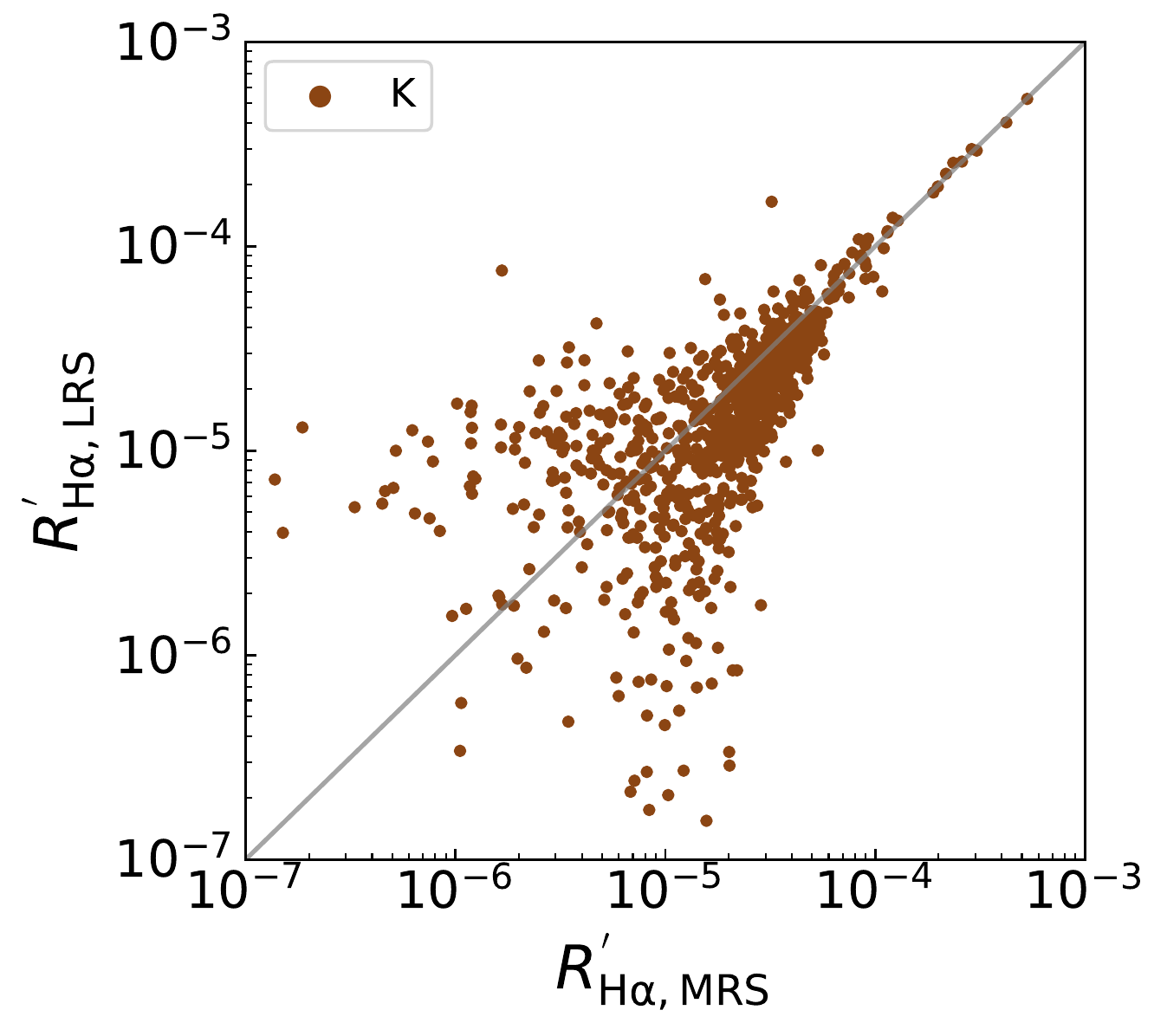}}
\centering
\subfigure[]{
\includegraphics[width=55ex]{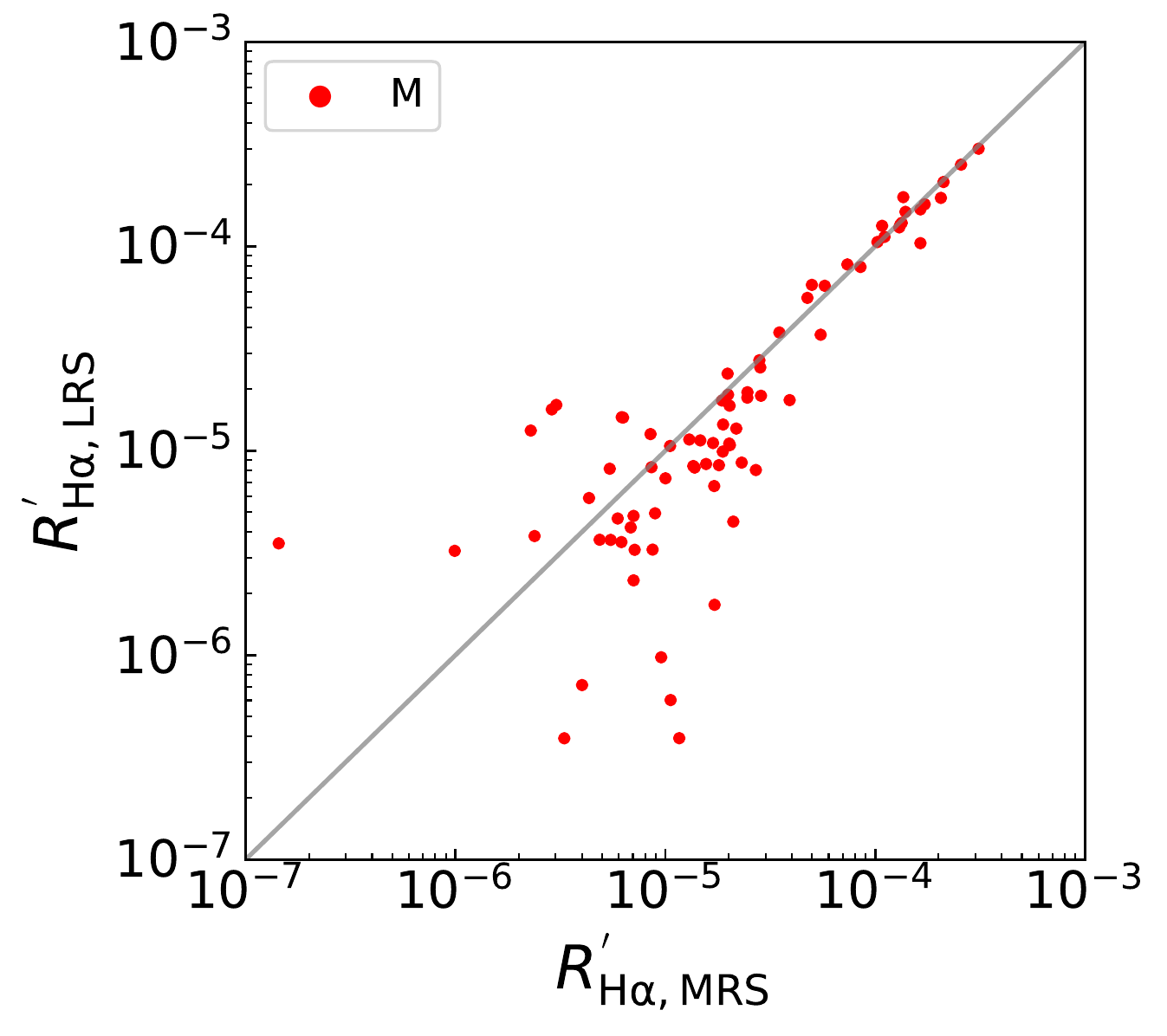}}
\caption{Comparisons of $R_{\rm{H\alpha}}^{'}$ derived from MRS and LRS observations for different types of stars.} 
\label{mrslrs.fig}
\end{figure*}

\begin{sidewaystable}
  \centering
    \caption{All the quantities calculated from the MRS data together with stellar parameters.}
    \label{tab:table3}
    \begin{tabular}{ccccccccccccc}
      \toprule 
       $\rm{ID}$ & 
       $EW^{'}$ &
       $\rm{log}(R_{\rm{H\alpha}}^{'})$ &
       $\Delta(\rm{log}(R_{\rm{H\alpha}}^{'}))$ &
       $EW^{+}$ &
       $\rm{log}(R_{\rm{H\alpha}}^{+})$ &
       $\Delta(\rm{log}(R_{\rm{H\alpha}}^{+}))$ & 
       $T_{\rm{eff}}$ &
       $\rm{log\emph{g}}$ & 
       $\rm{[Fe/H]}$ & 
       $\rm{log(\chi)}$ & 
       $\tau_{c}$ & 
       $P_{rot}$\\
       & (\AA) &  &  & (\AA) &  &  & (K) & (dex) & (dex) &  & (days)  & (days)\\
       (1) & (2) & (3) & (4) & (5) & (6) & (7) & (8) & (9) & (10) & (11) & (12) & (13) \\
      \hline \\ 
J105250.78+113922.2* & -0.36$\pm$0.03 & -4.4$\pm$-5.47 & 0.07 & -0.94$\pm$0.03 & -3.99$\pm$-5.47 & 0.03 & 6129.21 & 4.36 & -0.16 & -3.96 & 14.85 & 2.7\\
J105630.42+114349.8 & -0.11$\pm$0.16 & -4.95$\pm$-4.76 & 0.33 & -0.7$\pm$0.16 & -4.13$\pm$-4.76 & 0.1 & 6080.17 & 4.12 & -0.65 & -3.97 & - & -\\
J105637.75+113841.4 & -1.95$\pm$0.09 & -3.74$\pm$-5.09 & 0.03 & -2.42$\pm$0.09 & -3.64$\pm$-5.09 & 0.02 & 6610.14 & 4.17 & -1.98 & -4.03 & - & -\\
J035800.88+231205.2* & 0.2$\pm$0.02 & -4.67$\pm$-5.74 & 0.06 & -0.38$\pm$0.02 & -4.39$\pm$-5.74 & 0.04 & 6146.33 & 3.99 & -0.02 & -3.97 & 10.6 & 14.98\\
J105558.30+105838.5 & 0.07$\pm$0.03 & -5.14$\pm$-5.51 & 0.37 & -0.51$\pm$0.03 & -4.26$\pm$-5.51 & 0.04 & 6146.76 & 4.11 & -0.15 & -3.97 & - & -\\
J035935.33+231710.4 & -0.65$\pm$0.08 & -4.15$\pm$-5.05 & 0.07 & -1.25$\pm$0.08 & -3.86$\pm$-5.05 & 0.04 & 6029.56 & 4.4 & -0.04 & -3.96 & - & -\\
J105519.59+111046.1 & 0.11$\pm$0.02 & -4.93$\pm$-5.61 & 0.16 & -0.46$\pm$0.02 & -4.29$\pm$-5.61 & 0.03 & 6204.56 & 4.21 & -0.24 & -3.96 & - & -\\
... & ... & ... & ... & ... & ... & ... & ... & ...& ... & ... & ... & ...  \\
      \hline \\
    \end{tabular}
\footnotesize{(1) ID: Target ID. (2) $EW^{'}$: $EW$ of $\rm H_{\alpha}$ line, excluding the photospheric contribuiton. (3) $\rm{log}(R_{\rm{H\alpha}}^{'})$: Logarithmic $R_{\rm{H\alpha}}^{'}$. (4) $\Delta(\rm{log}(R_{\rm{H\alpha}}^{'}))$: Variability of logarithmic $R_{\rm{H\alpha}}^{'}$. (5) $EW^{+}$: $EW$ of $\rm H_{\alpha}$ line, excluding the photospheric contribuiton and chromospheric emission that are not related to magnetic activity. (6) $\rm{log}(R_{\rm{H\alpha}}^{+})$: Logarithmic $R_{\rm{H\alpha}}^{+}$. (7) $\Delta(\rm{log}(R_{\rm{H\alpha}}^{'}))$: Logarithmic variability of $R_{\rm{H\alpha}}^{+}$. (8) $T_{\rm{eff}}$: Effective temperature. (9) $\rm{log\emph{g}}$: Surface gravity. (10) $\rm{[Fe/H]}$: Metallicity. (11) $\rm{log(\chi)}$: Logarithmic $\rm{\chi}$. (12) $\tau_{c}$: Convective turnover time. (13) $P_{rot}$: Rotation period.}
\end{sidewaystable}

\begin{sidewaystable}
  \centering
    \caption{All the quantities calculated from the LRS data together with stellar parameters.}
    \label{tab:table4}
    \begin{tabular}{ccccccccccccc}
      \toprule 
       $\rm{ID}$ & 
       $EW^{'}$ &
       $\rm{log}(R_{\rm{H\alpha}}^{'})$ &
       $\Delta(\rm{log}(R_{\rm{H\alpha}}^{'}))$ &
       $EW^{+}$ &
       $\rm{log}(R_{\rm{H\alpha}}^{+})$ &
       $\Delta(\rm{log}(R_{\rm{H\alpha}}^{+}))$ & 
       $T_{\rm{eff}}$ &
       $\rm{log\emph{g}}$ & 
       $\rm{[Fe/H]}$ & 
       $\rm{log(\chi)}$ & 
       $\tau_{c}$ & 
       $P_{rot}$\\
       & (\AA) &  &  & (\AA) &  &  & (K) & (dex) & (dex) &  & (days)  & (days)\\
       (1) & (2) & (3) & (4) & (5) & (6) & (7) & (8) & (9) & (10) & (11) & (12) & (13) \\
       \hline \\
J105250.78+113922.2* & -0.43$\pm$0.01 & -4.33$\pm$-6.16 & 0.09 & -1.04$\pm$0.01 & -3.94$\pm$-6.16 & 0.04 & 6129.21 & 4.36 & -0.16 & -3.96 & 14.85 & 2.7\\
J105630.42+114349.8 & -0.08$\pm$0.12 & -5.08$\pm$-4.9 & 0.58 & -0.7$\pm$0.12 & -4.13$\pm$-4.9 & 0.21 & 6080.17 & 4.12 & -0.65 & -3.97 & - & -\\
J105637.75+113841.4 & -1.88$\pm$0.05 & -3.75$\pm$-5.34 & 0.04 & -2.42$\pm$0.05 & -3.64$\pm$-5.34 & 0.03 & 6610.14 & 4.17 & -1.98 & -4.03 & - & -\\
J105613.73+111240.5 & -0.22$\pm$0.1 & -4.6$\pm$-4.93 & 0.12 & -0.84$\pm$0.1 & -4.02$\pm$-4.93 & 0.28 & 6079.54 & 4.24 & -0.15 & -3.95 & - & -\\
J035800.88+231205.2* & 0.05$\pm$0.0 & -5.27$\pm$-6.36 & 0.4 & -0.56$\pm$0.0 & -4.22$\pm$-6.36 & 0.05 & 6146.33 & 3.99 & -0.02 & -3.97 & 10.6 & 14.98\\
J105558.30+105838.5 & -0.04$\pm$0.01 & -5.38$\pm$-6.13 & 0.39 & -0.65$\pm$0.01 & -4.15$\pm$-6.13 & 0.03 & 6146.76 & 4.11 & -0.15 & -3.97 & - & -\\
J035935.33+231710.4 & -0.78$\pm$0.04 & -4.07$\pm$-5.4 & 0.07 & -1.41$\pm$0.04 & -3.81$\pm$-5.4 & 0.04 & 6029.56 & 4.4 & -0.04 & -3.96 & - & -\\
... & ... & ... & ... & ... & ... & ... & ... & ...& ... & ... & ... & ...  \\
      \hline \\
    \end{tabular}
\footnotesize{(1) ID: Target ID. (2) $EW^{'}$: $EW$ of $\rm H_{\alpha}$ line, excluding the photospheric contribution. (3) $\rm{log}(R_{\rm{H\alpha}}^{'})$: Logarithmic $R_{\rm{H\alpha}}^{'}$. (4) $\Delta(\rm{log}(R_{\rm{H\alpha}}^{'}))$: Variability of logarithmic $R_{\rm{H\alpha}}^{'}$. (5) $EW^{+}$: $EW$ of $\rm H_{\alpha}$ line, excluding the photospheric contribution and chromospheric emission that are not related to magnetic activity. (6) $\rm{log}(R_{\rm{H\alpha}}^{+}$: Logarithmic $R_{\rm{H\alpha}}^{+}$. (7) $\Delta(\rm{log}(R_{\rm{H\alpha}}^{'}))$: Logarithmic variability of $R_{\rm{H\alpha}}^{+}$. (8) $T_{\rm{eff}}$: Effective temperature. (9) $\rm{log\emph{g}}$: Surface gravity. (10) $\rm{[Fe/H]}$: Metallicity. (11) $\rm{log(\chi)}$: Logarithmic $\rm{\chi}$. (12) $\tau_{c}$: Convective turnover time. (13) $P_{rot}$: Rotation period.}
\end{sidewaystable}

\subsection{$\chi$ and $R_{\rm H_{\rm \alpha}}^{'}$}

\cite{2004PASP..116.1105W} proposed a distance-independent method to calculate the normalized luminosity of $\rm{H_{\alpha}}$ lines: $L_{ \rm H_{\alpha}} / L_{\rm bol}$, or $R_{\rm H_{\rm \alpha}}^{'}$:
\begin{equation}
R_{\rm H_{\rm \alpha}}^{'} = L_{\rm H_{\alpha}} / L_{\rm bol} = \chi \times EW^{'},
\end{equation}
In this work, $\chi$ was estimated as
following:
\begin{equation}
\chi = \frac{f_{\rm \lambda 6564}}{f_{\rm bol}} = \frac{f_{\rm \lambda 6564}}{\sigma T_{\rm eff}^{4}}.
\end{equation}

The continuum flux $f_{\rm \lambda 6564}$ was estimated based on the \emph{PHOENIX} synthetic spectra. We fitted the continuum of these spectra and used the flux at 6564 \AA\ as $f_{\rm \lambda 6564}$. Figure \ref{chi.fig} shows a clear trend of $\chi$ as a function of effective temperature. For each target we calculated a median value of $R_{\rm{H\alpha}}^{'}$ using multiple observations.
The errors of $R_{\rm{H\alpha}}^{'}$ were calculated following error propagation and the median value of the errors in multiple observations was used for each target.

\begin{figure*}
\centering
\subfigure[]{
\includegraphics[width=0.45\textwidth]{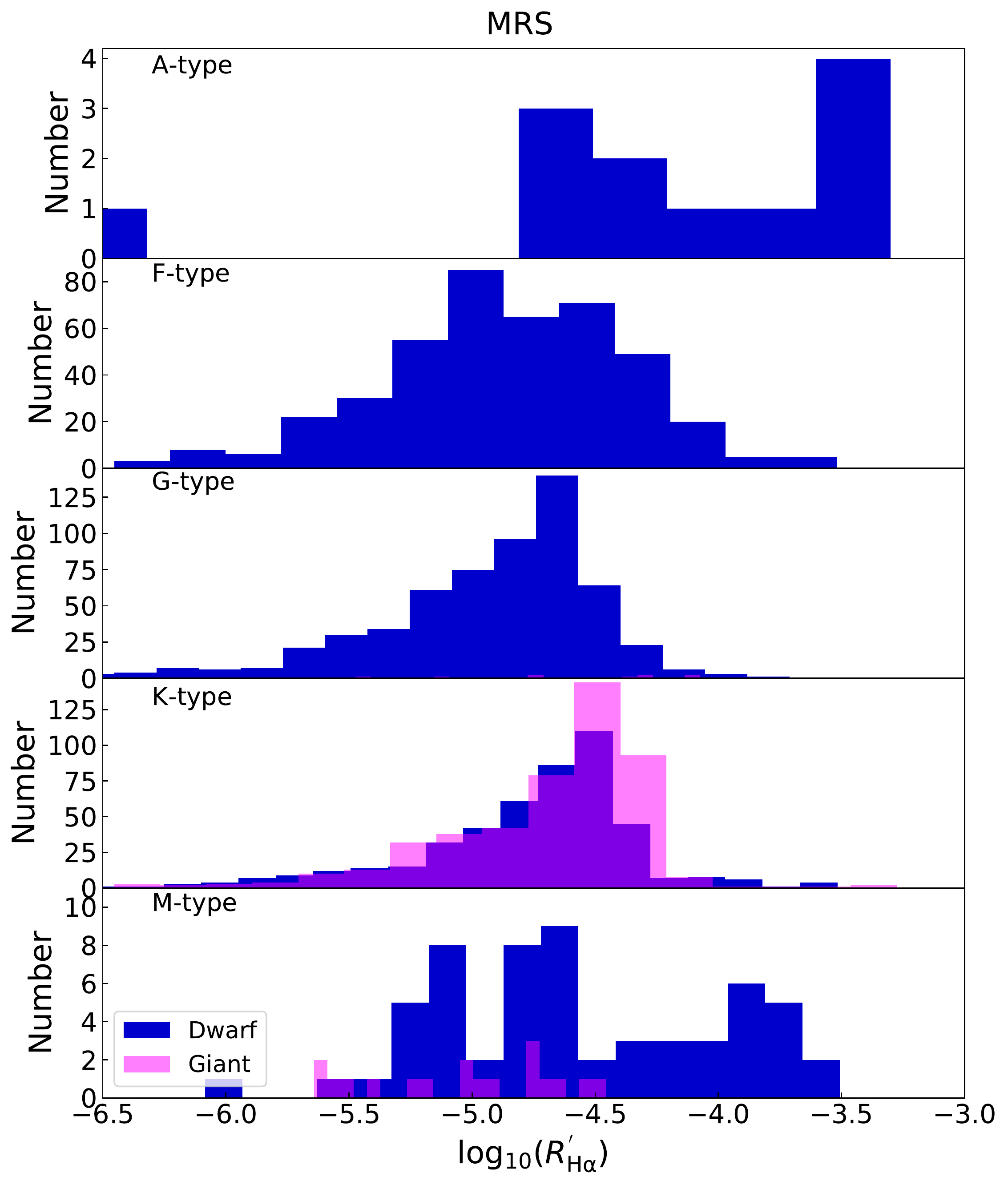}}
\subfigure[]{
\includegraphics[width=0.45\textwidth]{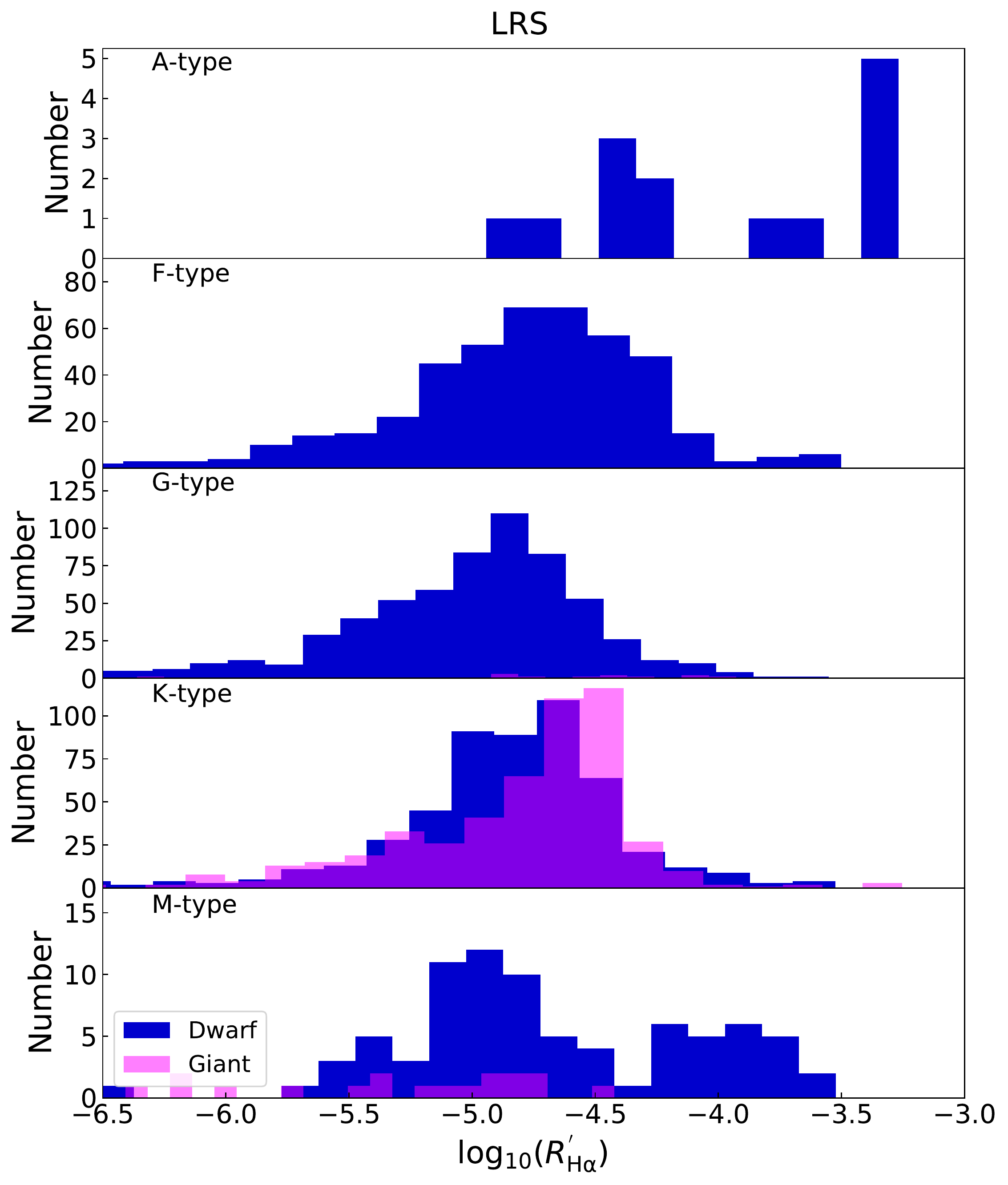}}
\caption{Histograms of stellar activity levels. Dwarfs and giants are denoted by blue and purple colours, respectively.}
\label{hist.fig}
\end{figure*}

The comparisons of $R_{\rm H_{\rm \alpha}}^{'}$ between the MRS and LRS observations are shown in Figure \ref{mrslrs.fig}. Mostly, the results from the two datasets are in good agreement. Figure \ref{hist.fig} shows the distribution of $R_{\rm H_{\rm \alpha}}^{'}$ for different types of stars. 
These stars share a similar range of emission levels, with $R_{\rm H_{\rm \alpha}}^{'}$ ranging from $-$5 to $-$3.5.
M giants tend to have weaker $\rm{H_{\alpha}}$ emission compared to M dwarfs while the $R_{\rm H_{\rm \alpha}}^{'}$ values are similar for K dwarfs and giants.

\subsection{Rotation periods and Rossby number}
As described in Section \ref{sel.sec}, the \emph{Lomb-Scargle} method was used to determine the period from the $K$2 data, and the folded light curves were visually checked and the \emph{Kepler Data Integration Platform}\footnote{http://kepler.bao.ac.cn} \citep{2019ApJS..241...29Y} was adopted to improve our efficiency. 
For A-type stars in our sample, we did not detect reliable rotation period, while for 36 F-type stars, we detected rotational modulations in their light curves. Rotational modulations have been found in the light curves of many hot stars, which can be explained to be caused by starspots or other co-rotating structures \citep[e.g.][]{2011MNRAS.415.1691B, 2019MNRAS.485.3457B}. However, it is debated that the rotational modulation for early-type stars may have non-magnetic origins \citep{2020MNRAS.497.4117L, 2020MNRAS.498.2456S}. For example, the $g$ modes, which are excited by resonant couplings, can be presented in many early-type main-sequence stars. These modes could result in frequencies that are consistent with those of photometric rotational modulations and harmonics \citep{2020MNRAS.497.4117L}. In our sample, only nine early-type F stars ($T_{\rm eff} > $ 6500 K) exhibit rotational periods, six of which have $R_{\rm H_{\rm \alpha}}^{'}$ measurements. Performing new observations (such as high-resolution spectroscopic monitoring) capable of detecting or ruling out the presence of starspots for these early-F stars is beyond the scope of this work.

\begin{figure}
\centering
\includegraphics[width=0.45\textwidth]{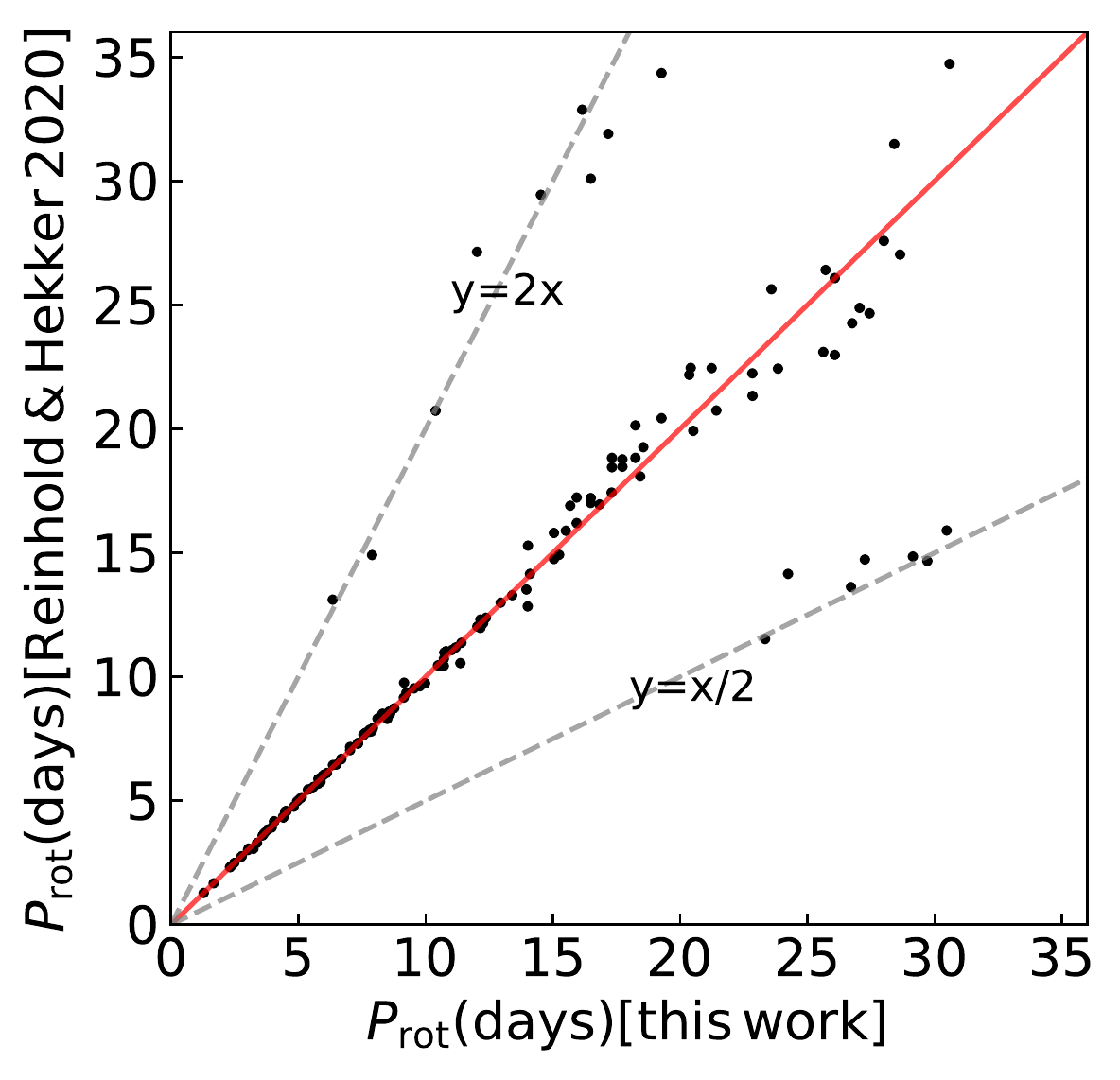}
\caption{Comparison of rotation periods derived in this work and in \cite{2020A&A...635A..43R}. }
\label{rot.fig}
\end{figure}

Totally 296 targets with accurate period measurements were picked out. Among them there are 226 targets with measured MRS $R_{\rm{H\alpha}}^{'}$ and 242 targets with measured LRS $R_{\rm{H\alpha}}^{'}$, which are marked with ``*" in Table \ref{tab:table3} and Table \ref{tab:table4}, respectively. \cite{2020A&A...635A..43R} also measured rotation periods of \emph{K2} targets. To check the accuracy of our results, we compared the rotation periods of common targets between this work and \cite{2020A&A...635A..43R}. Most of the rotation periods are in good agreement (Figure \ref{rot.fig}). The objects with large deviations mostly have period ratio being 1/2 or 2, indicating that the period measurements from \cite{2020A&A...635A..43R} are half or double of the period of our work. In addition, the rotation periods of F-type stars in this work are consistent with those from \cite{2020A&A...635A..43R}. 


In this work the grid models of stellar evolution tracks from the Yale-Potsdam Stellar Isochrones (YAPSI) were used to calculate the convective turnover time $\tau_{c}$ \citep{2017ApJ...838..161S}. We adopted all the sub-grids with $Y = 0.28$, which contains solar calibration and includes different metallicity grids of $[\rm{Fe/H}]$ of $+$0.3, 0.0, $-$0.5, $-$1.0, and $-$1.5. 
We derived the location of each star in the $T_{\rm{eff}}$--$\rm{log}\emph{g}$ diagram and compared it with these model evolutionary tracks. The model points located inside the parameter uncertainties were selected to calculate a median value of $\tau_{c}$. We repeated this step for all metallicity grids and the final $\tau_{c}$ of each target was estimated through an interpolation of $\tau_{c}$ among these metallicity grids.
Note that none of the early-F stars has $\tau_{c}$ estimation, suggesting that they won't affect following analysis on the activity-rotation relation.

Finally the Rossby number was calculated as Ro $= P_{\rm rot}$/$\tau_{c}$. There are 195 targets with estimations of both Rossby number and MRS $R_{\rm{H\alpha}}^{'}$, and 203 targets with estimations of both Rossby number and LRS $R_{\rm{H\alpha}}^{'}$.
In Table \ref{tab:table3} and Table \ref{tab:table4} we list all the results of our samples, including the median values of $EW^{'}$ and $R_{\rm{H\alpha}}^{'}$, the stellar parameters ($T_{\rm eff}$, log$g$ and [Fe/H]), rotational period ($P_{\rm rot}$) and Rossby number (Ro), etc. 

\section{Discussion}
\subsection{$R_{\rm{H\alpha}}^{'}$ and Rossby number}
\label{relation.sec}

Stellar rotation plays a key role in generating magnetic fields. The relations between different activity proxies and rotation have been extensively investigated \citep[e.g.,][]{2003A&A...397..147P,2011ApJ...743...48W,2014ApJ...795..161D, 2017ApJ...834...85N}. 
Activity-rotation relation is usually suggested to consist of two distinct sequences: the saturated region for rapidly rotating stars, in which the activity level is constant, and the power-law decay region for slowly rotating stars, where the activity level is rotation-dependent.

\begin{figure*}
\centering
\includegraphics[width=0.96\textwidth]{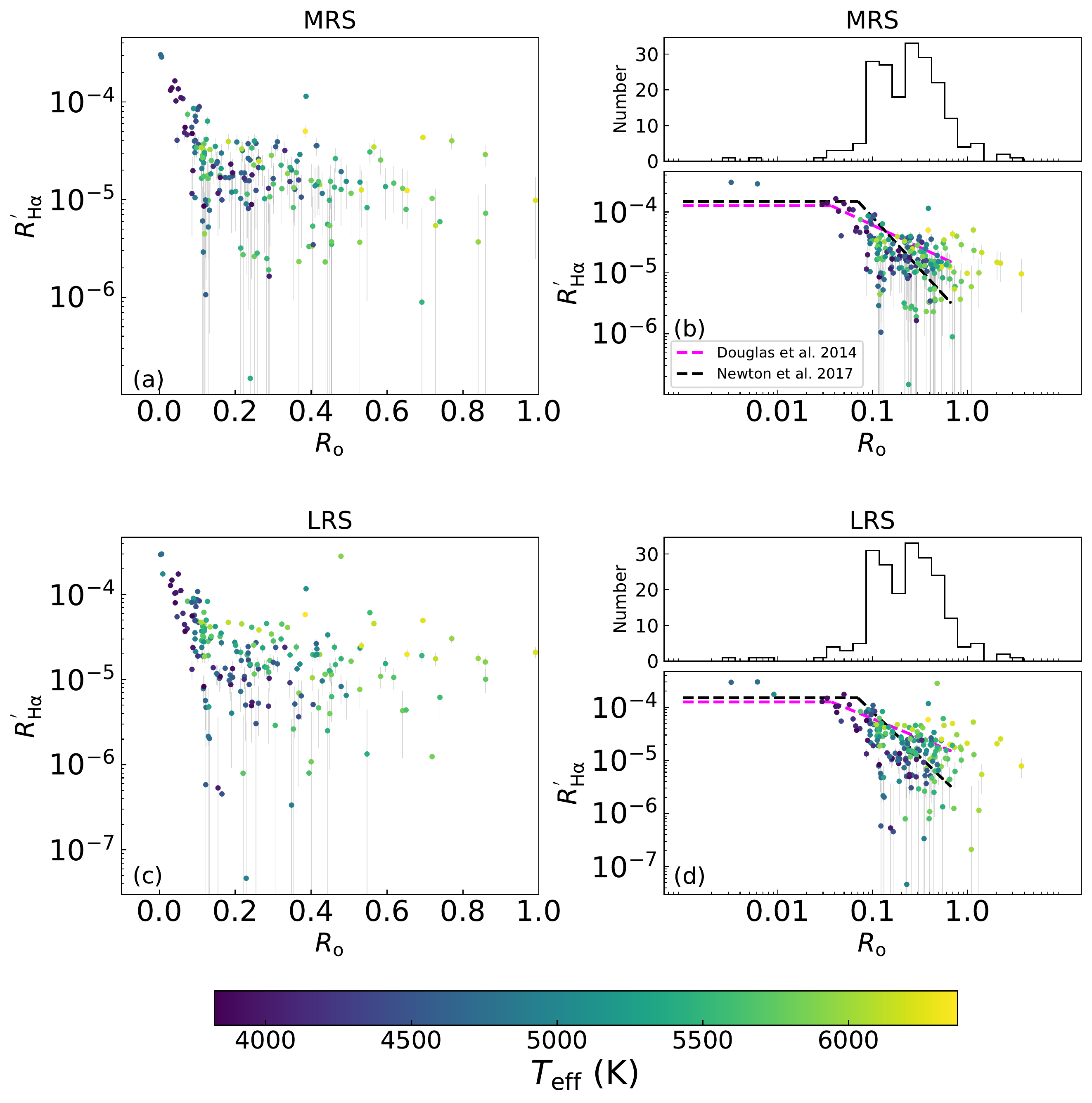}
\caption{$R_{\rm{H\alpha}}^{'}$ - $R_{\rm{o}}$ relations for MRS and LRS observations. In panel (a) and panel (c) the relations are in linear-log scale while in panel (b) and panel (d) are in log-log scale. Points with different colours represent different effective temperatures. Errors are shown in grey lines. Dashed magenta and black lines are the activity-rotation relations from \cite{2014ApJ...795..161D} and \cite{2017ApJ...834...85N}, respectively. Both the relations were shifted by Ro$/$3. Histograms along the $R_{\rm{o}}$ axis are also presented.}
\label{relation1.fig}
\end{figure*}

However, some basic issues on this relation are still under debate:
(1) Whether the activity level in the saturation region keeps increasing or constant when rotation velocity increases?
(2) Where does the transition from the saturation to the non-saturation region occur?
(3) Whether the non-saturation region follows a single power-law?


Some previous studies have suggested a slight slope of the activity-rotation relation in the saturated region, indicating a remaining dependence of the activity levels on rotation periods even for active, fast-rotating stars \citep[e.g.][]{2008ApJ...687.1264M,2014ApJ...794..144R}. 
By using the dwarfs observed by both \emph{Kepler} and \emph{XMM-Newton}, \cite{2019A&A...628A..41P} found that some objects, especially F-type stars, clearly deviate from the standard decay power-law relation.
In addition, \cite{2018A&A...618A..48M} revised the relation based on different activity indicators including X-rays, Ca \scriptsize{\uppercase\expandafter{\romannumeral2}} \normalsize H$\&$K and $\rm{H_{\alpha}}$ emissions. They divided the relation into four regions: a saturated region, a fast decay region, and two slowly decay regions with different power-law shapes. The results of these studies are quite different from the standard picture \citep{2003A&A...397..147P,2011ApJ...743...48W}.


\begin{figure*}
\centering
\subfigure[]{
\includegraphics[width=0.45\textwidth]{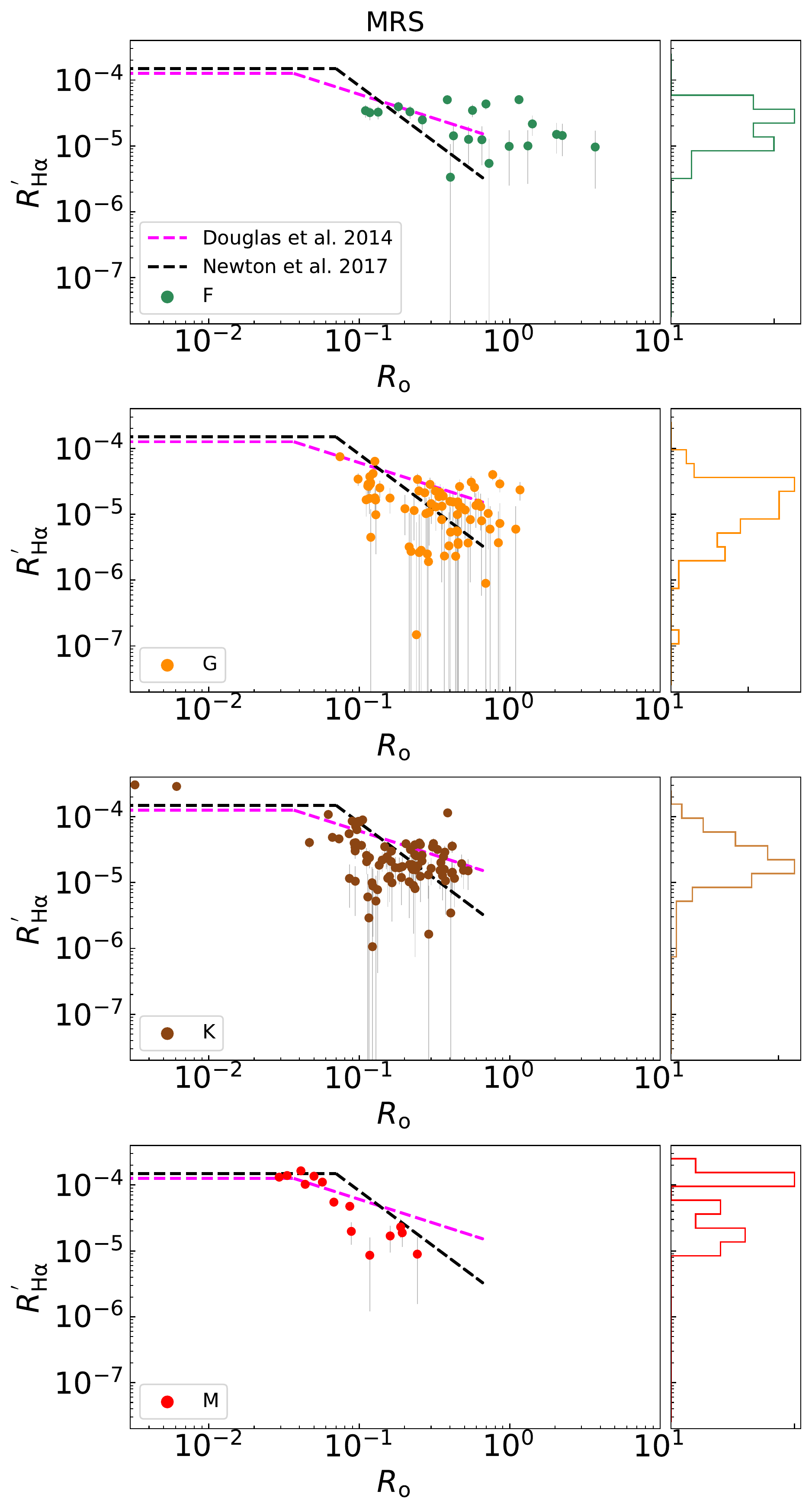}}
\subfigure[]{
\includegraphics[width=0.45\textwidth]{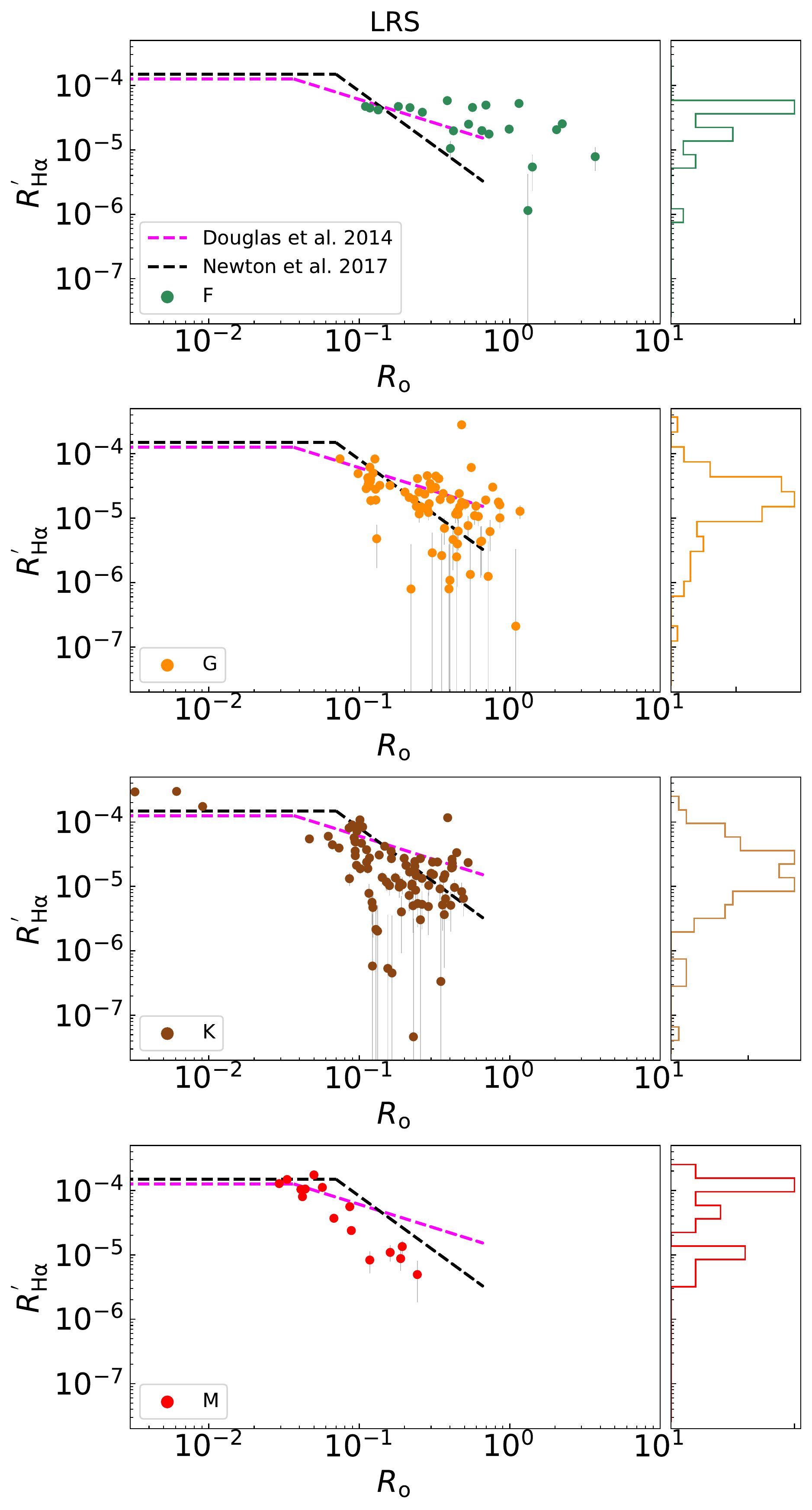}}
\caption{Activity-rotation relations of different kinds of stars. From top to bottom corresponds to F-, G-, K-, and M-type stars. Panel (a) shows the results of MRS data and panel (b) displays the results of LRS data. Again the dashed magenta and black lines are shifted relation of \cite{2014ApJ...795..161D} and \cite{2017ApJ...834...85N}, respectively. Histograms along the $R_{\rm{H\alpha}}^{'}$ are also shown.}
\label{relation2.fig}
\end{figure*}

Figure \ref{relation1.fig} displays the $R_{\rm{H\alpha}}^{'}$-Ro relations for both the MRS and LRS data. No giant star is plotted since there are only five giants with well-defined Rossby number in our sample. Our results suggest a complicated profile for the activity-rotation relation. In the saturated region, our targets agree with the standard model from previous literature (Figure \ref{relation1.fig}), suggesting that as the Rossby number decreases, the activity level would keep the same. However, only a few targets in our sample lie in this region, more sources are required to confirm whether there is a slight slope \citep[e.g.][]{2008ApJ...687.1264M,2014ApJ...794..144R}.

The knee points seem to be varying among different types of stars (Figure \ref{relation2.fig}). In previous studies, the knee point that separates the saturated and non-saturated regions is at Ro $=$ 0.13 \citep{2011ApJ...743...48W}. 
Alternatively, \cite{2018A&A...618A..48M} argued that the activity level would keep unchanged only when Ro $<$ 0.021, while \cite{2017ApJ...834...85N} showed that the break occurs near Ro $=$ 0.2. Note that for these studies, Ro values were calculated from the classical empirical estimation of $\tau$, which is about 1/3 of the theoretical values of $\tau_{c}$ in our work \citep{2020NatAs...4..658L,2020ApJ...902..114W}.
It seems that there are more than one knee point in the activity-rotation relation, or different types of stars may have different knee points (Figure \ref{relation2.fig}).
However, due to the limited number of targets in the saturated region, we cannot give an accurate estimation of the positions of knee points.

Our results suggested that the fast decay region clearly cannot be fit by a single power-law. Stars with different spectral types exhibit different slopes (Figure \ref{relation2.fig}), and a mix of them would lead to a messy relation in the decay region (Figure \ref{relation1.fig}). For example, the slope of F-type stars is gentler than that of cooler stars. In the same range of Ro, hot stars (e.g., F stars) seem to be more active in ${\rm H_{\alpha}}$ emission than cool stars (e.g., M stars). Similar phenomenon has been found in activity-rotation relations constructed by different activity proxies. As the temperature increases, the slope of the decay region gradually changes, indicating that such tendency is universal \citep{2019A&A...628A..41P, 2020A&A...635A..43R}.



\subsection{Basal flux and $R_{\rm{H\alpha}}^{+}$}

\cite{1987A&A...172..111S} pointed out that the basal fluxes of chromospheric lines (e.g., Ca \scriptsize{\uppercase\expandafter{\romannumeral2}} \normalsize H$\&$K) represent the non-radiative heating in the outer atmosphere, which is unrelated to magnetic activity.
The dynamo-driven magnetic activity can be calculated as the excess flux above the baseline, which can be constructed with inactive stars.
Here we further studied the impact of such baseline on stellar activity.

The baseline was fitted based on the most inactive stars in our sample (Figure \ref{basal.fig}).
Then the pure chromospheric emission due to magnetic activity of $\rm{H_{\alpha}}$ line was defined as the chromospheric flux excess following \cite{2013A&A...549A.117M, 2018A&A...618A..48M},
\begin{equation}
\begin{split}
EW^{+}  &= EW' - EW_{\rm{basal}}\\
&=EW - EW_{\rm phot} - EW_{\rm{basal}}.
\end{split}
\end{equation}
Then $EW^{+}$ was converted to $R_{\rm{H\alpha}}^{+}$ with the $\chi$ following
\begin{equation}
R_{\rm H_{\rm \alpha}}^{+} = L_{\rm H_{\alpha}} / L_{\rm bol} = \chi \times EW^{+}.
\end{equation}
All the $EW^{+}$ and $R_{\rm{H\alpha}}^{+}$ values are listed in Table \ref{tab:table3} and \ref{tab:table4}. 

\begin{figure*}
\centering
\includegraphics[width=0.96\textwidth]{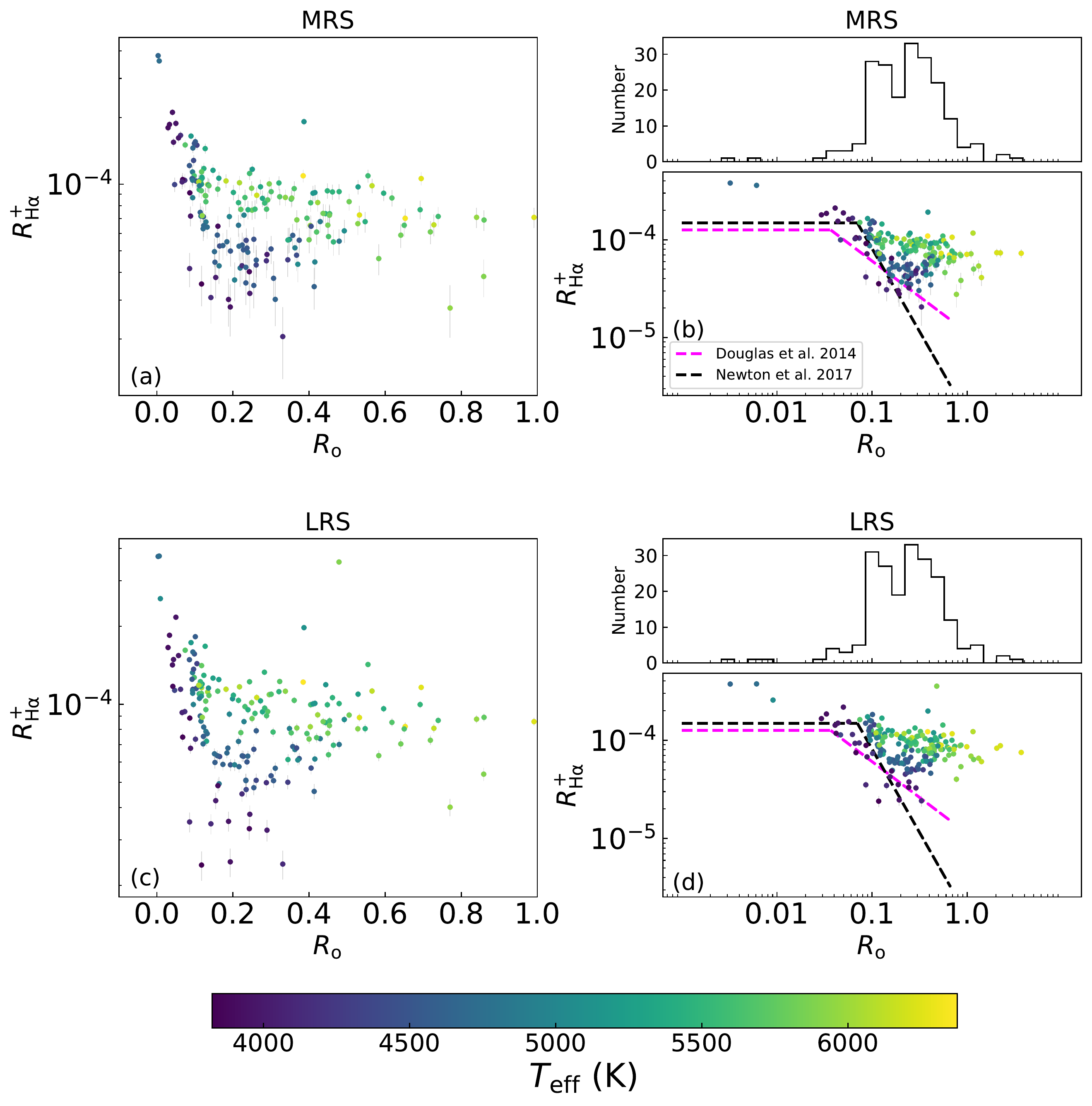}
\caption{Same as Figure \ref{relation1.fig} but for $R_{\rm{H\alpha}}^{+}$.}
\label{relation_base.fig}
\end{figure*}

\begin{figure*}
\centering
\subfigure[]{
\includegraphics[width=0.45\textwidth]{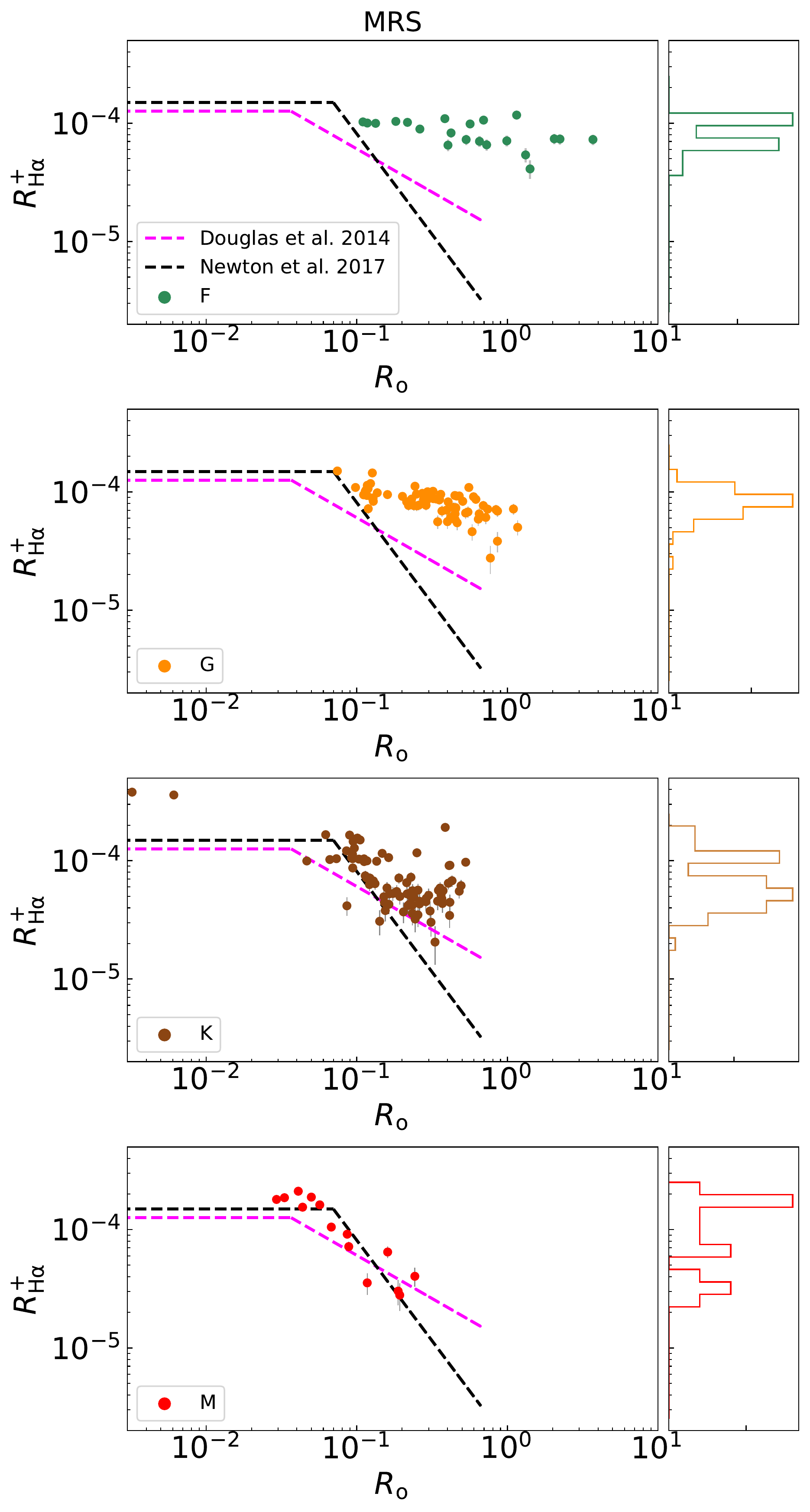}}
\subfigure[]{
\includegraphics[width=0.45\textwidth]{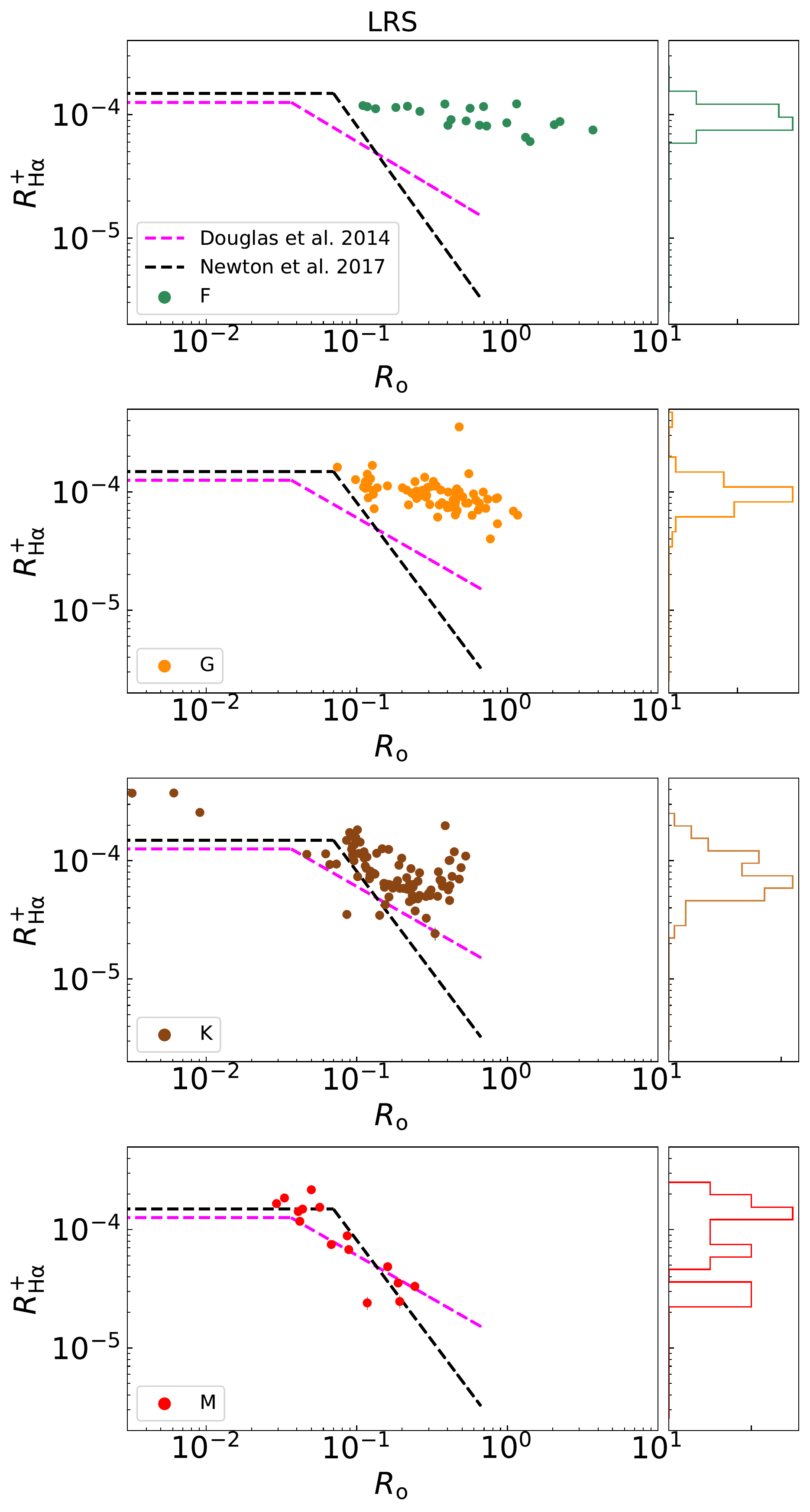}}
\caption{Same as Figure \ref{relation2.fig} but for $R_{\rm{H\alpha}}^{+}$.}
\label{relation3.fig}
\end{figure*}

The $R_{\rm{H\alpha}}^{+}$-$R_{o}$ relations (Figure \ref{relation_base.fig} and \ref{relation3.fig}) show some differences compared to the $R_{\rm{H\alpha}}^{'}$-$R_{o}$ relations.
The subtraction of basal flux increases the activity levels of all stars. Now the hot stars and cool stars seem to be divided into two separated groups.
Hot stars show much flatter slope and higher activity in the decay region.
The $R_{\rm{H\alpha}}^{+}$-$R_{o}$ relations of F- and G-type stars significantly deviate from the relations of \cite{2014ApJ...795..161D} and \cite{2017ApJ...834...85N}.
This again indicates the complex profile of the decay region, and it cannot be fitted by a single power-law. 

Although the relations from \cite{2017ApJ...834...85N} and \cite{2014ApJ...795..161D} were constructed from M stars, the deviation of hot stars is still worth being detailed investigated.
As the temperature increases, the decay slope also gradually becomes flat. Meanwhile, 
the activity levels of hot stars increase more after the subtraction of basal flux.
It's hard to tell whether this behaviour is real or it's a fake due to the poor fitting of basal flux.
However, as shown in previous study, hot stars do exhibit higher X-ray activity level than cool stars, although their differences are small \citep{2019A&A...628A..41P}.

\begin{figure*}
\centering
\includegraphics[width=0.96\textwidth]{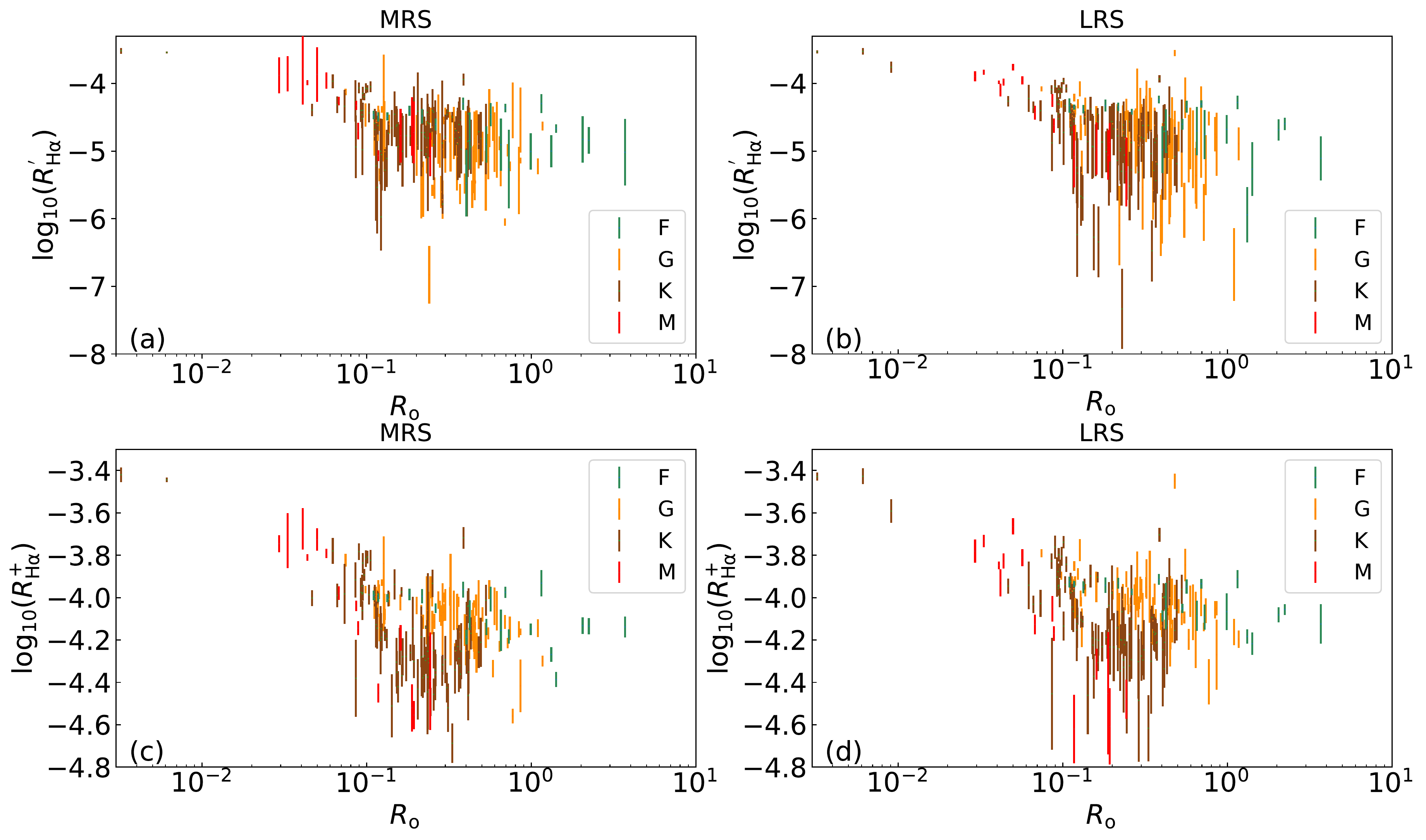}
\caption{Variability of the activity levels for different types of stars. The length of lines represent the variation ranges of $R_{\rm{H\alpha}}^{'}$ (upper panels) and $R_{\rm{H\alpha}}^{+}$ (lower panels) for each target, based on their multiple observations.}
\label{relation1_var.fig}
\end{figure*}

One possible origin of the large scatters of stellar activity in the decay region (shown in Figure \ref{relation2.fig} and \ref{relation3.fig}) may be the variation of the magnetic activity. As shown in \citet{2020ApJ...902..114W}, the variation of stellar X-ray activity is universal and significant. 
Figure \ref{relation1_var.fig} displays clear variability of ${\rm{H\alpha}}$ emissions for our sample stars, suggesting notable stellar chromospheric activity variations.


\subsection{$R_{\rm{H\alpha}}^{'}$ and Light Curves}

\begin{figure*}
\centering
\subfigure[]{
\includegraphics[width=0.45\textwidth]{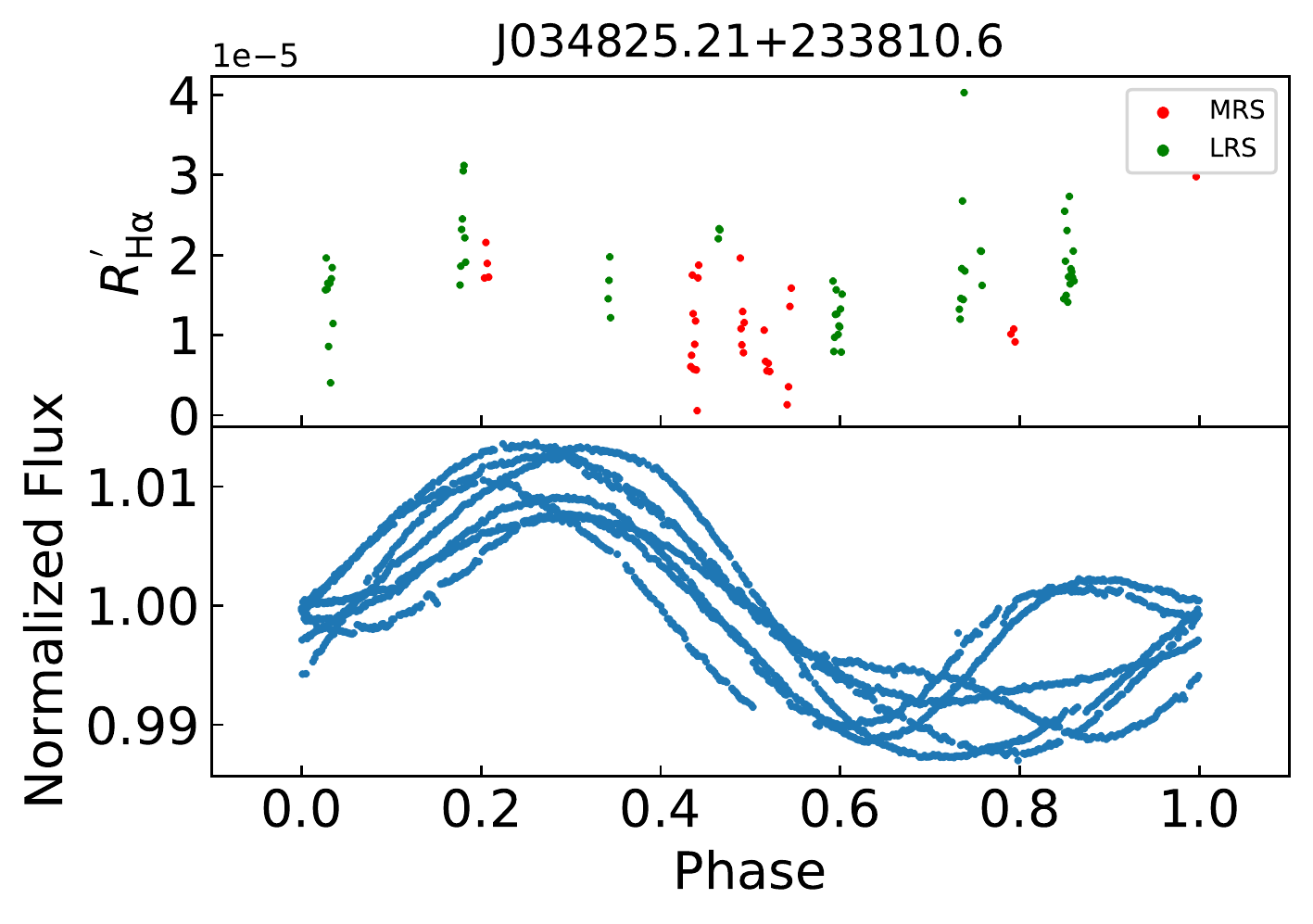}}
\subfigure[]{
\includegraphics[width=0.45\textwidth]{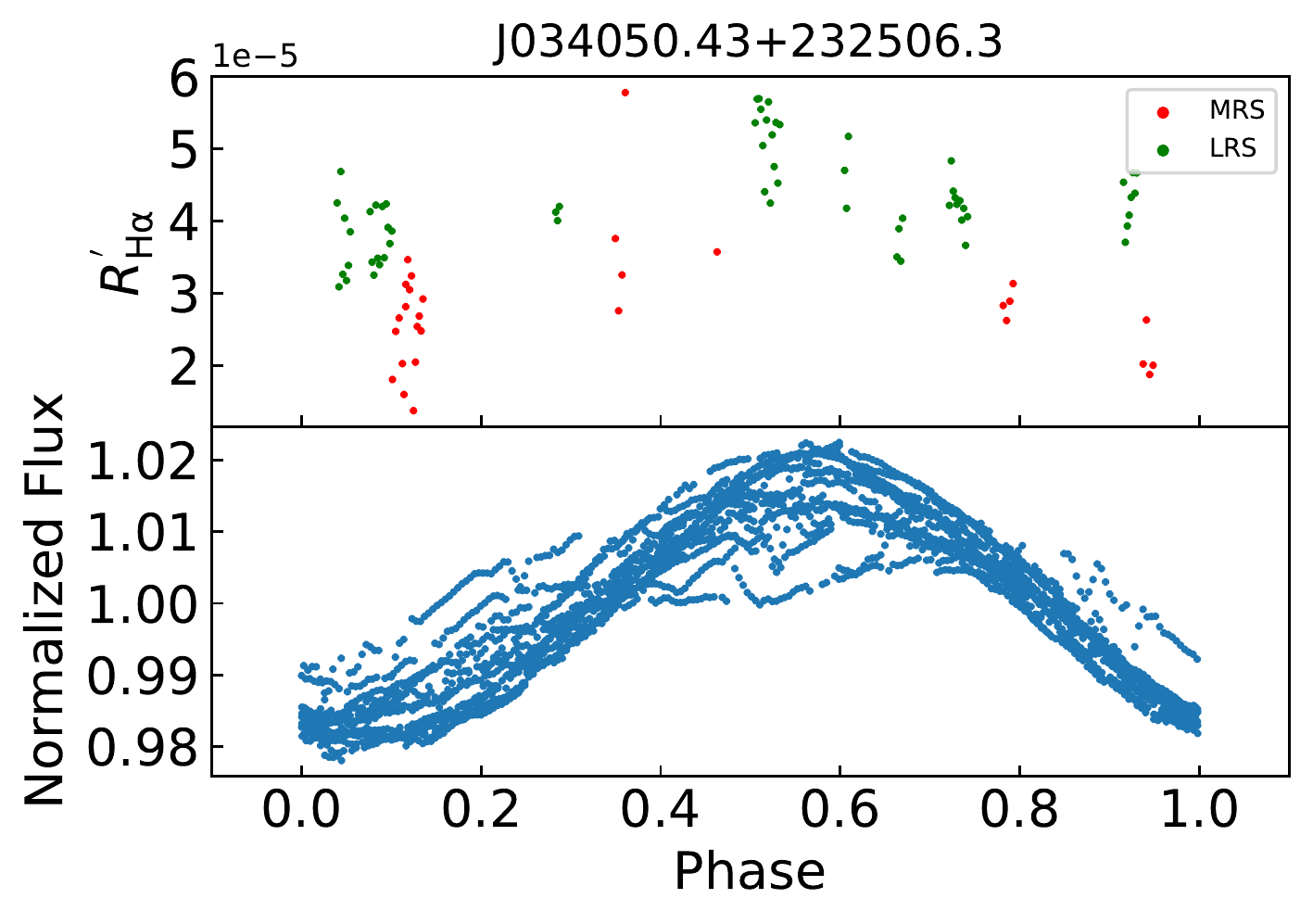}}
\subfigure[]{
\includegraphics[width=0.45\textwidth]{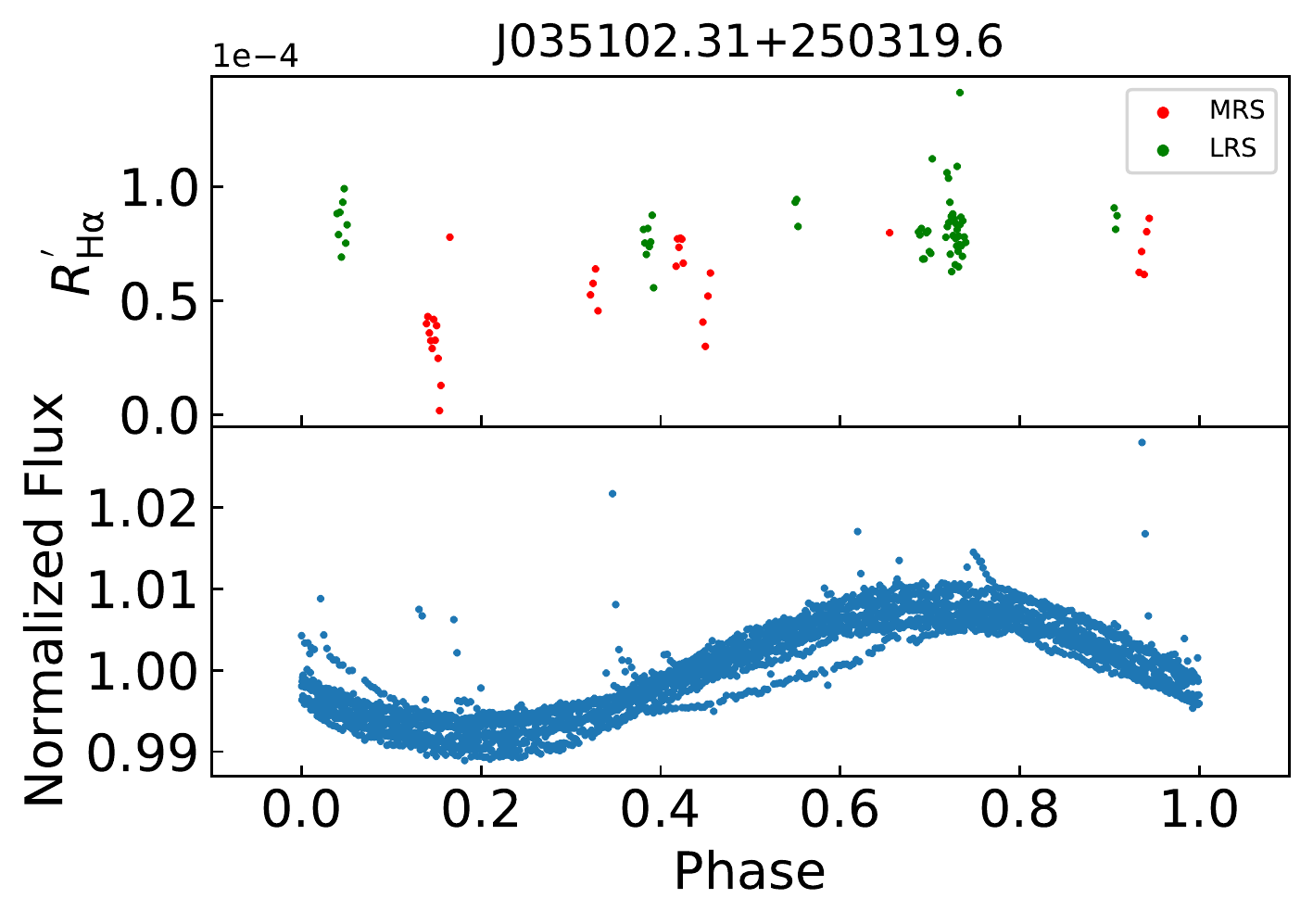}}
\caption{Plots of both phase folded light curves and $R_{\rm{H\alpha}}^{'}$ series.}
\label{lc.fig}
\end{figure*}


Amplitudes of light curves could be modulated by (cool) starspots or (hot) faculae. The photometric activity proxy $R_{\rm{eff}}$, which is used to quantified sinusoidal modulation of light curves, is proportional to the classical chromospheric proxy $R_{\rm{HK}}^{'}$ \citep{2020ApJS..247....9Z}. 
It is known that large solar plages are spatially associated with sunspots \citep{2017ApJ...835..158M}.
Some previous studies also showed that there was an anti-correlation in the rotational phase between
the chromospheric activity proxies (e.g., Ca \scriptsize{\uppercase\expandafter{\romannumeral2}} \normalsize H$\&$K, $\rm{H_{\alpha}}$) and light curves: the line emission was stronger when the amplitude of light curve was near minimum \citep{1982AJ.....87.1546D,1999A&A...349..863B,2010RAA....10..253F,2020MNRAS.495.2949F}.
This indicates that the plages are located in the outer chromosphere overlying the starspots in the visible photosphere.  

By using the LAMOST time-domain spectra and \emph{K2} light curves, we also investigated relations between $R_{\rm{H\alpha}}^{'}$ and rotational phase of light curves. For targets with well detected rotation periods, both the light curves and $R_{\rm{H\alpha}}^{'}$ values of multiple observations were folded by their rotation periods. 
We found that for the vast majority of targets, $R_{\rm{H\alpha}}^{'}$ and optical light curves have no obvious correlation. Only three targets show possible correlation (Figure \ref{lc.fig}), J034825.21+233810.6 (EPIC 211041648), J034050.43+232506.3 (EPIC 211028209) and J035102.31+250319.6 (EPIC 211129308) their $R_{\rm{H\alpha}}^{'}$ series correlates well with the light curve. Meanwhile, same targets would be remained after changing the $R_{\rm{H\alpha}}^{'}$ to $R_{\rm{H\alpha}}^{+}$. 

The positive correlation between the ${\rm{H\alpha}}$ emission and photospheric variability suggests that the later is mainly dominated by faculae instead of starspots. It is possible since \citet{2017NatAs...1..612S} presented that the contribution of faculae to the variability of Total Solar irradiance is comparable to that of spots at timescales from 2 to 7 days. \cite{2020A&A...635A..43R} showed a gap (mainly around 10 to 20 days) in the rotation period distribution and interpreted it as a cancellation between bright faculae and dark spots. The gap is consistent with the period range of 15--25 days which implies a transition from spot-dominated to faculae-dominated activity \citep{2017ApJ...851..116M}.


The correlation between $R_{\rm{H\alpha}}^{'}$ and optical light curves may be coincidental, since their observations are not simultaneous. 
The evolution timescale of starspots or faculae may be extremely shorter than the spectroscopic and photometric observation intervals.
These may explain why no clear correlation was found for most targets.
However, we found that the folded light curves of some targets  change slightly during the four-year observation, indicating that the active regions evolve slowly. 
The three objects also show stable light curves (Figure \ref{lc.fig}), suggesting that the light curves can be used to compare with the LAMOST observations, even if the photometric and spectroscopic observations were not simultaneous.
Future simultaneous photometric and spectroscopic observations for more targets could shed more light on this issue.

\section{Summary}

In this work, we systematically studied the statistical properties of $\rm{H_{\alpha}}$ lines based on the LAMOST TD spectra. 
The chromospheric emission was estimated through two steps: (1) Subtracting the photospheric contribution from the observed $\rm{H_{\alpha}}$ lines, i,e., $EW^{'} = EW - EW_{\rm phot}$. An index of $R_{\rm{H\alpha}}^{'}$ was calculated from $EW^{'}$. (2) Besides the photospheric part, a baseline, which was thought to be unrelated to chromorpheric heating,
was fitted from inactive stars and further subtracted, i.e., $EW^{+} = EW - EW_{\rm phot} - EW_{\rm{basal}}$. This leads to an estimation of another index $R_{\rm{H\alpha}}^{+}$.

Both the $R_{\rm{H\alpha}}^{'}$-$R_{o}$ relation and $R_{\rm{H\alpha}}^{+}$-$R_{o}$ relation were investigated.
Besides the typical divided saturation region and non-saturation region, they both show complicated profiles in the latter decay regime.
Hot stars show flatter slopes and higher activity level than cool stars. Such phenomenon is more notable after the baseline line was subtracted. 
This suggests that different stars may follow different power laws in the decay region.
Alternatively, this may be caused by the larger variability of ${\rm{H\alpha}}$ emission, which was revealed by multiple observations.
In addition, the differences between the $R_{\rm{H\alpha}}^{'}$-$R_{o}$ relation and $R_{\rm{H\alpha}}^{+}$-$R_{o}$ relation tells the sensitivity of these indices to the selection of basal flux. The fitting and subtraction of the baseline strongly affects the distribution of activity levels and the activity-rotation relations of different types of stars, which should be carefully studied for the chromospheric activity proxies.

By using the TD photometric and spectroscopic data, we also investigated the phased variations of $\rm{H_{\alpha}}$ emission and optical light curves.
Only three targets exhibit positive correlations, indicating their light curves are dominated by hot faculae. 
Further simultaneous photometric and spectroscopic observations will be a key to study this correlation.

\acknowledgements
We thank the anonymous referee for helpful comments and suggestions that have significantly improved the paper.
We acknowledge the support from National Key Research and Development Program of China (NKRDPC) under grant numbers 2019YFA0405000 and 2019YFA0405504. 
This work is also supported by National Science Foundation of China (NSFC) under grant numbers 11988101/11933004.
We also thank the support from Strategic Priority Program of the Chinese Academy of Sciences under grant number XDB41000000.
Henggeng Han acknowledges the support from CAS-DAAD Joint Fellowship Programme for Doctoral students of UCAS.
Song Wang acknowledges the support from the Youth Innovation Promotion Association of the CAS (id. 2019057).

\bibliographystyle{aasjournal}
\bibliography{main}

\clearpage
\appendix
\renewcommand\thefigure{\Alph{section}\arabic{figure}}
\section{Calculating the $EWs$ with fixed and narrow wavelength range}

In order to test whether the chromospheric activities would mainly contribute to the line core of $\rm{H_{\alpha}}$, which is similar to the situation of Ca \scriptsize{\uppercase\expandafter{\romannumeral2}} \normalsize H$\&$K lines, we repeated the calculation processes but the $EWs$ of $\rm{H_{\alpha}}$ lines was computed with a fixed integration interval (i.e., 10 \AA) for all kinds of stars. Such a narrow integration interval could also avoid the contamination coming from possible blending lines.

\setcounter{figure}{0}
\begin{figure}[h]
\centering
\subfigure[]{
\includegraphics[width=0.45\textwidth]{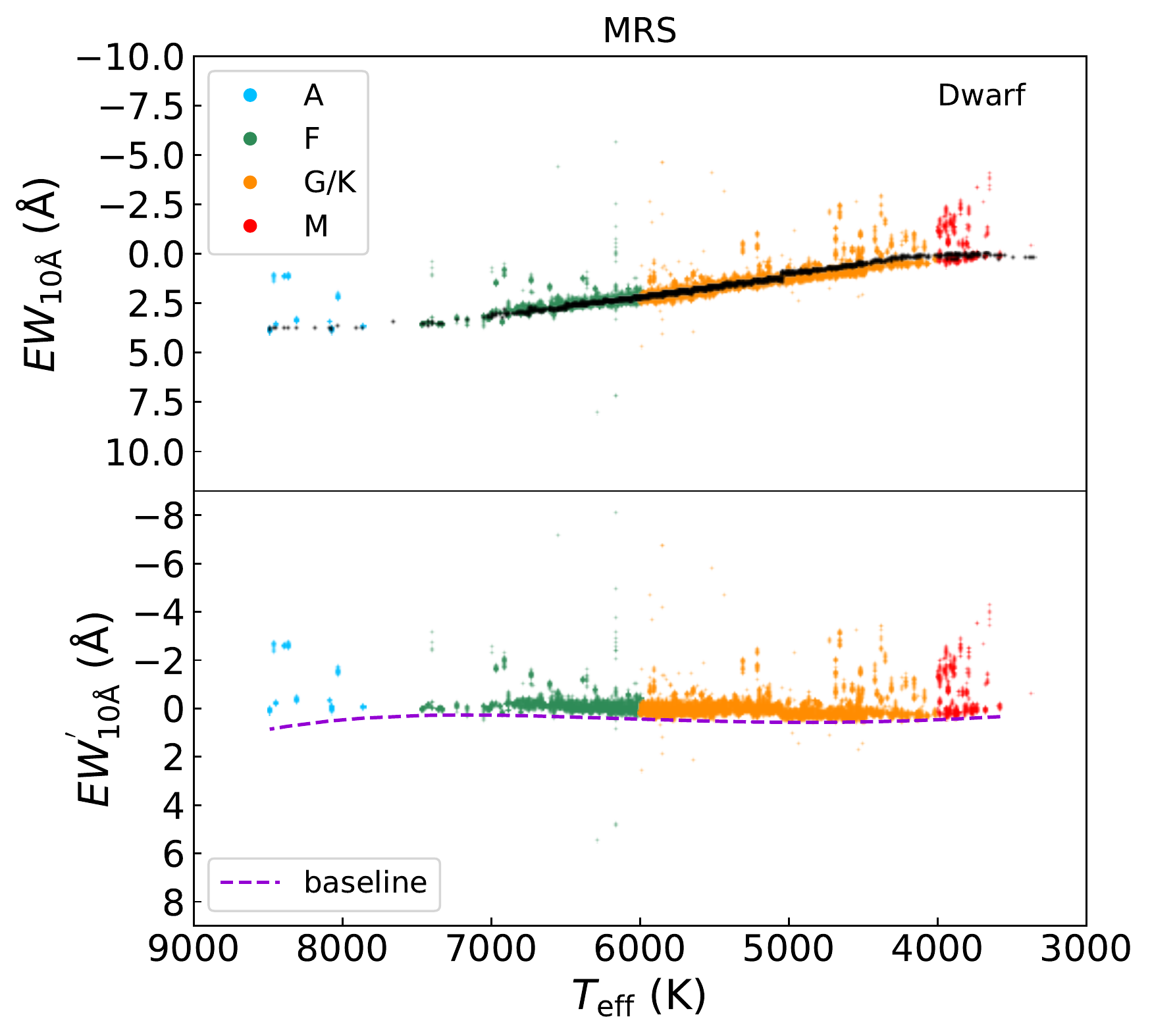}}
\subfigure[]{
\includegraphics[width=0.45\textwidth]{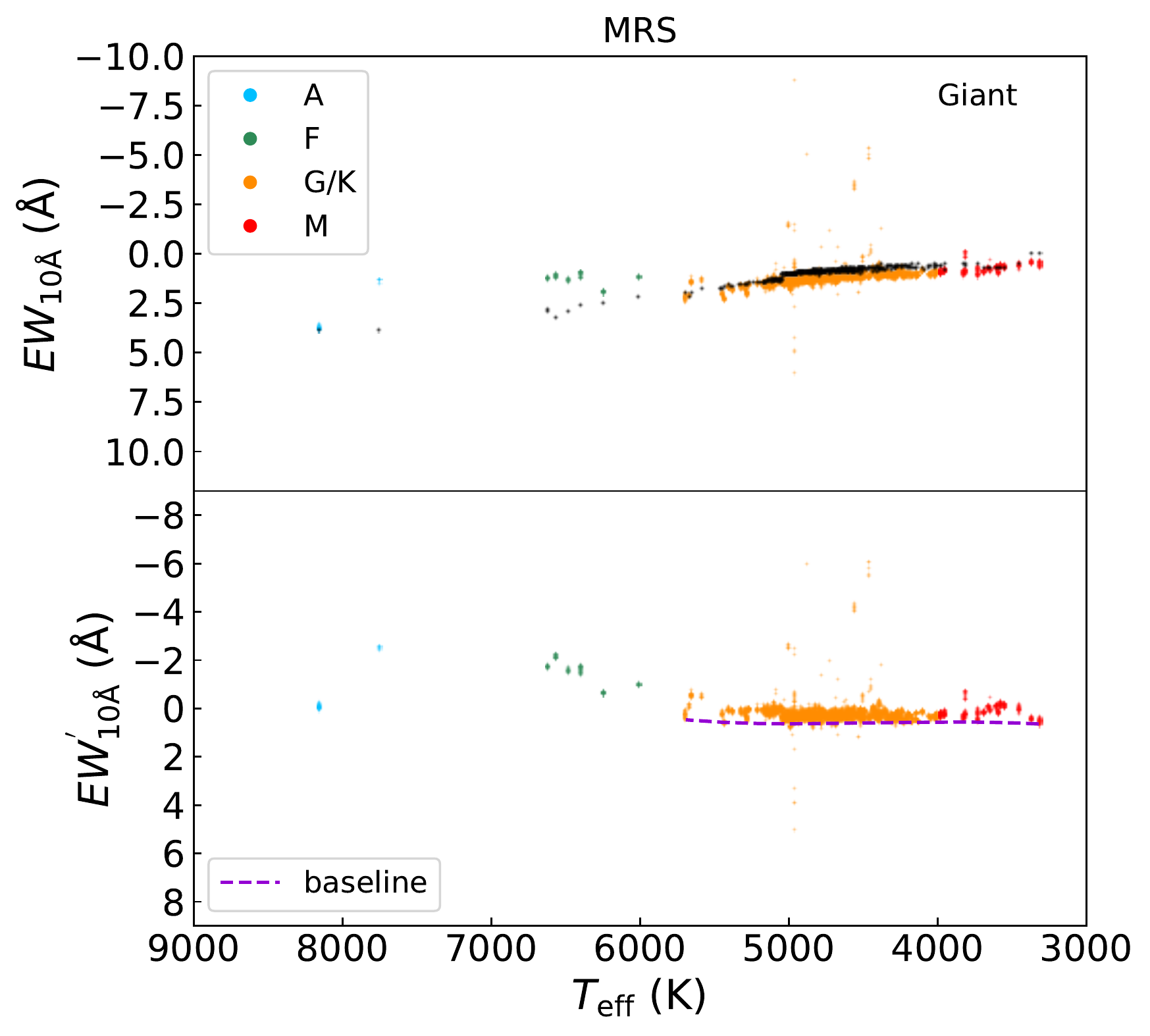}}

\caption{Same as Figure \ref{basal.fig} but for integration interval of 10\AA.}

\end{figure}

Same as Figure \ref{basal.fig}, in Figure A1 we plot the $EWs$ against effective temperatures. Panel (a) and panel (b) show the results of dwarf stars and giants, respectively. Different colours represent different kinds of stars. Meanwhile, black dots represent photospheric contribution to $\rm{H_{\alpha}}$ lines, which were also derived based on the \emph{PHOENIX} synthetic spectra \citep{2013A&A...553A...6H}. But this time, the wavelength interval for the integration was also 10\AA. 

\setcounter{figure}{1}
\begin{figure}[h]
\centering
\includegraphics[width=0.45\textwidth]{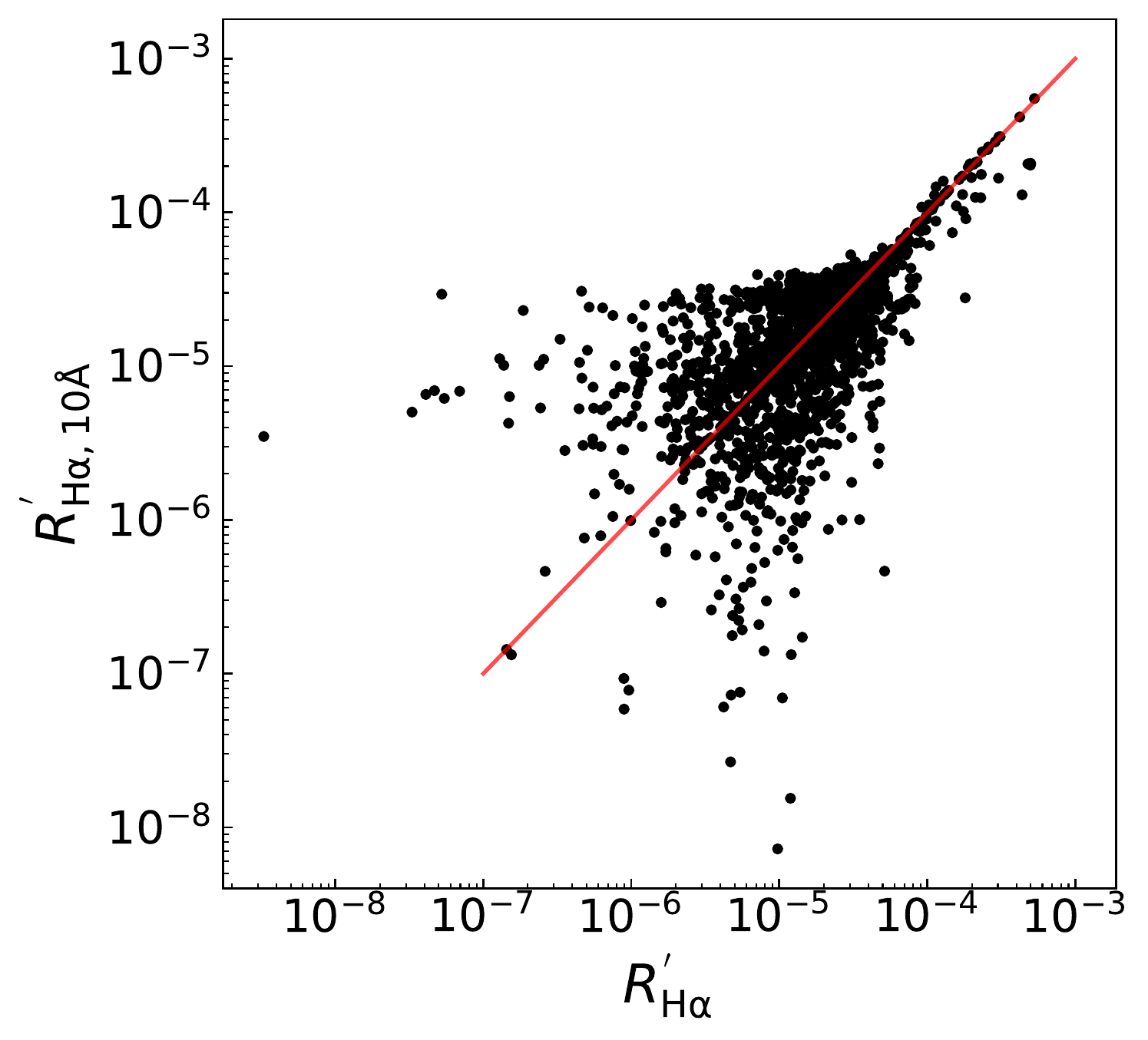}
\caption{Comparison between $R_{\rm{H\alpha}, 10 \AA}^{'}$ and $R_{\rm{H\alpha}}^{'}$.}

\end{figure}

Then the $EW$s were corrected using Equation (2) and further converted to normalized luminosity $R_{\rm{H\alpha}}^{'}$ based on Equation (3), in which the $\chi$ were from section 3.3. We compared the newly derived $R_{\rm{H\alpha}}^{'}$, named, $R_{\rm{H\alpha}, 10 \AA}^{'}$ and those calculated from various integration intervals. Obviously, for most of the targets $R_{\rm{H\alpha}, 10 \AA}^{'}$ agree well with the $R_{\rm{H\alpha}}^{'}$(Figure A2). 


\setcounter{figure}{2}
\begin{figure}[h]
\centering
\includegraphics[width=110ex]{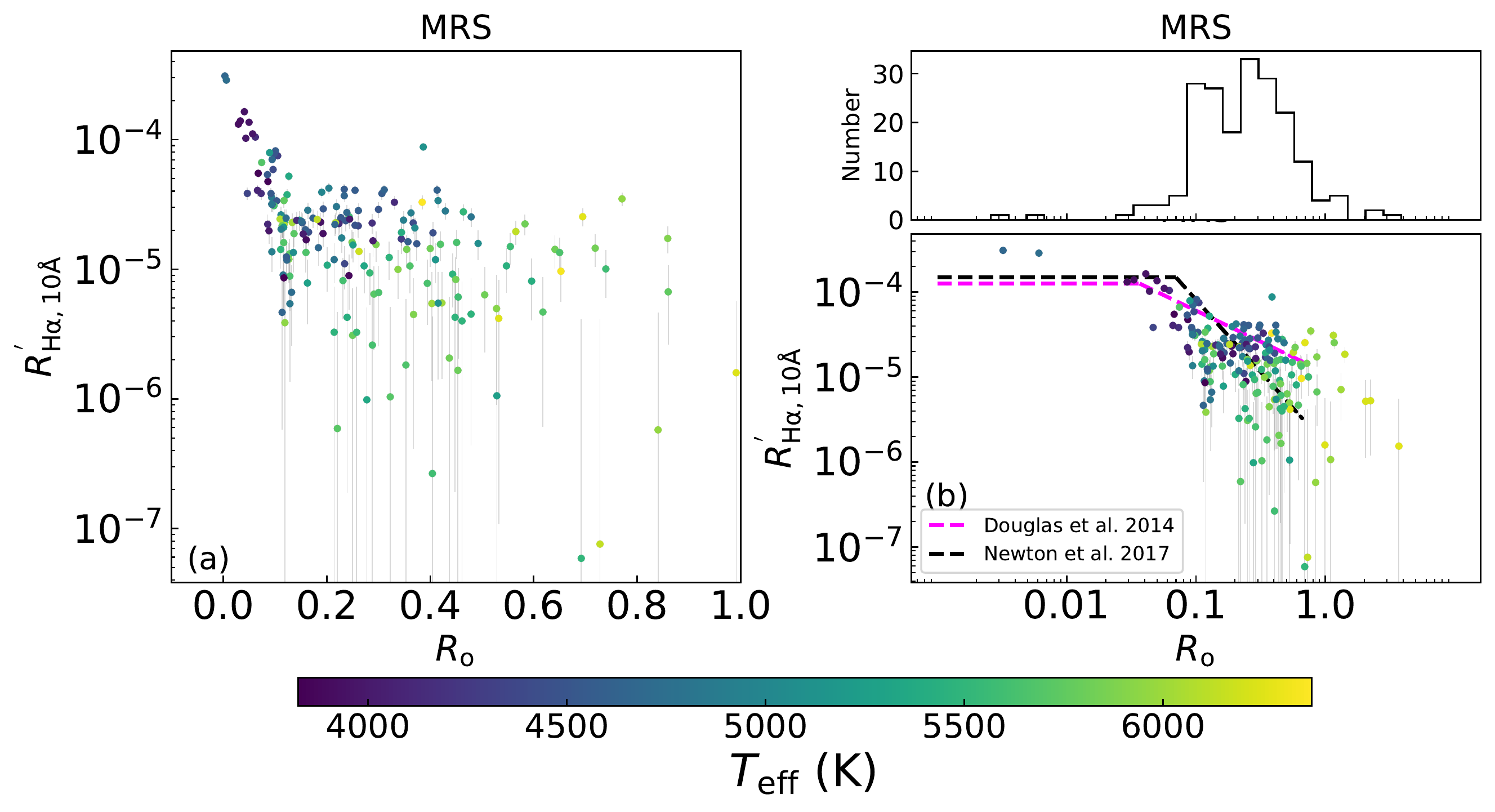}
\caption{Activity-rotations based on $R_{\rm{H\alpha}, 10 \AA}^{'}$. Same as Figure \ref{relation1.fig}, the plot is in linear-log scale in panel (a) and in log-log scale in panel (b).}

\end{figure}

Meanwhile, the activity-rotation relations were then renewed based on $R_{\rm{H\alpha}, 10 \AA}^{'}$. Same as Figure \ref{relation1.fig}, we plot the relation in different scales, i.e., linear-log scale in panel (a) and log-log scale in panel (b) of Figure A3. 
Dashed black and magenta lines are relations from \cite{2014ApJ...795..161D} and \cite{2017ApJ...834...85N}, respectively, which were also shifted by Ro$/$3. Apparently, the $R_{\rm{H\alpha}, 10 \AA}^{'}$-Ro relation is similar to the $R_{\rm{H\alpha}}^{'}$-Ro relation. 
In the non-saturated region the relation cannot be described by a simple power-law. Different stars exhibit different slopes in the decay region.

Furthermore, the $R_{\rm{H\alpha}, 10 \AA}^{'}$ were converted to $R_{\rm{H\alpha}, 10 \AA}^{+}$ through subtracting the basal fluxes and the $R_{\rm{H\alpha, 10 \AA}}^{+}$-Ro relation is given in Figure A4. It is clear that hot stars then tend to exhibit higher activity levels compared to cool stars. The F- and G-type stars deviate significantly from the relations given by \cite{2017ApJ...834...85N} and \cite{2014ApJ...795..161D}, suggesting that the selection of basal flux could strongly affect the activity-relations .


\setcounter{figure}{3}
\begin{figure}[h]
\centering
\includegraphics[width=110ex]{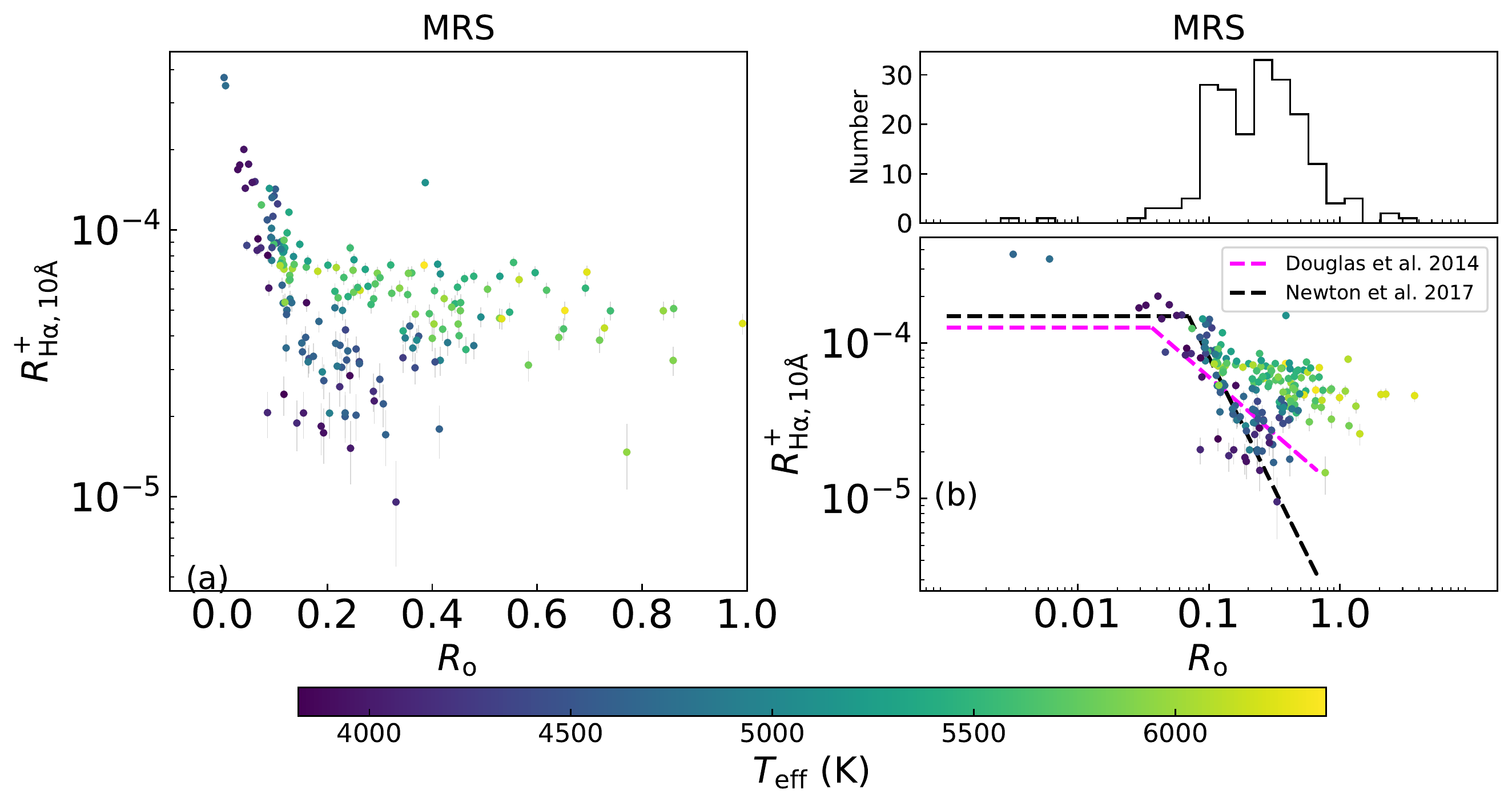}
\caption{Same as Figure \ref{relation1.fig} but for $R_{\rm{H\alpha}, 10 \AA}^{+}$}

\end{figure}

\end{CJK*}
\end{document}